\documentclass[twocolumn,trackchanges]{aastex62}
\usepackage{graphicx}
\usepackage{amsmath}
\usepackage{amssymb}
\usepackage{wasysym}

\shorttitle{The sub-Neptune desert}
\shortauthors{McDonald et al.}

\begin{document}


\title{The sub-Neptune desert and its dependence on stellar type: Controlled by lifetime X-ray irradiation}
\date{\today}

\author{George D. McDonald}
\correspondingauthor{George D. McDonald}
\email{g.mcdonald@rutgers.edu}
\altaffiliation{E-mail as of: \today}
\affiliation{School of Earth \& Atmospheric Sciences, Georgia Institute of Technology, 311 Ferst Drive, Atlanta, GA 30332, USA}

\author{Laura Kreidberg}
\affiliation{Harvard-Smithsonian Center for Astrophysics, 60 Garden Street, Cambridge, MA 02138, USA}
\affiliation{Harvard Society of Fellows, 78 Mount Auburn Street, Cambridge, MA 02138, USA}

\author{Eric Lopez}
\affiliation{NASA Goddard Space Flight Center, 8800 Greenbelt Road, Greenbelt, MD 20771, USA}
\affiliation{GSFC Sellers Exoplanet Environments Collaboration, NASA GSFC, Greenbelt, MD 20771}

\begin{abstract}
Short-period sub-Neptunes with substantial volatile envelopes are among the most common type of known exoplanets. However, recent studies of the \textit{Kepler} population have suggested a dearth of sub-Neptunes on highly irradiated orbits, where they are vulnerable to atmospheric photoevaporation. Physically, we expect this ``photoevaporation desert'' to depend on the total lifetime X-ray and extreme ultraviolet flux, the main drivers of atmospheric escape. In this work, we study the demographics of sub-Neptunes as a function of lifetime exposure to high energy radiation and host star mass. We find that for a given present day insolation, planets orbiting a 0.3 $M_{\astrosun}$ star experience $\sim$100 $\times$ more X-ray flux over their lifetimes versus a 1.2 $M_{\astrosun}$ star. Defining the photoevaporation desert as a region consistent with zero occurrence at 2 $\sigma$, the onset of the desert happens for integrated X-ray fluxes greater than 1.43 $\times 10^{22}$ erg/cm$^2$ to 8.23 $\times 10^{20}$ erg/cm$^2$ as a function of planetary radii for 1.8 -- 4 $R_{\oplus}$. We also compare the location of the photoevaporation desert for different stellar types. We find much greater variability in the desert onset in bolometric flux space compared to integrated X-ray flux space, suggestive of photoevaporation driven by steady state stellar X-ray emissions as the dominant control on desert location. Finally, we report tentative evidence for the sub-Neptune valley, first seen around Sun-like stars, for M \& K dwarfs. The discovery of additional planets around low-mass stars from surveys such as the TESS mission will enable detailed exploration of these trends.
\end{abstract}

\section{Introduction}
The \textit{Kepler} mission has confirmed the existence of over 2341 extrasolar planets \citep{Thompson2017}. These discoveries provide valuable information about the distribution of planets in terms of their radii and semi-major axes. Of particular significance has been the discovery that given the current sample of close-in planets (5-50 day periods), the most common type are planets with radii in the range of 1 -- 1.5 or 2 -- 2.8 $R_{\oplus}$ (Earth radii) \citep{Fressin2013, Petigura2013,Petigura2013b,Fulton2017}. The former are typically referred to as super-Earths, we hereafter refer to the latter, as well as all planets in the range 1.8 $< R_{\oplus} <$ 4, as sub-Neptunes.

There is substantial evidence that these sub-Neptune planets represent a distinct population. Their boundary on the low radius end lies with evidence that planets with $R_{\oplus} >$ 1.5 are not purely rocky in composition \citep{Weiss2014,Rogers2015}, with those with $R_{\oplus} >$ 1.7 in particular likely to require significant volatile envelopes to explain their low densities \citep[e.g.,][]{Lopez2014, Rogers2015,Fulton2017}. Subsequent studies have shown that these planets are best explained by models that have hydrogen/helium (hereafter H/He) envelopes atop Earth-like rocky cores \citep[e.g.,][]{Wolfgang2015,Chen2016,Lopez2016a,Owen2017,Jin2017}. The upper bound of $<$ 4 $R_{\oplus}$ is motivated by the fact that these planets are typically significantly more massive \citep{Weiss2014,Wolfgang2016}, which makes these planets considerably more resistant than smaller planets to atmospheric photoevaporation from high-energy radiation \citep{Lopez2013}. Despite their commonalities, studies of the mass-radius relations of these sub-Neptunes indicate that they are a diverse population, spanning a wide range of bulk densities and compositions \citep[][e.g.,]{Weiss2014,Wolfgang2016}. Understanding the set of conditions that will determine the envelope compositions and radii of these sub-Neptunes is a major area of research and will be key to our understanding of planet formation. Furthermore, it is important to recognize that the observed densities and radii are snapshots in time, and that an evolution across different portions of the parameter space is probable for many sub-Neptunes.

A specific process expected to have dramatic consequences on the evolution and properties of the sub-Neptune population is photoevaporation driven by high energy radiation, which has been suggested as a mechanism that would strip sub-Neptunes of their H/He envelopes over time \citep[e.g.,][]{Baraffe2006,Jackson2010,Owen2012, Lopez2012, Owen2013,Chen2016}. This process is most impactful for sub-Neptunes on short-period orbits, where short means $\lesssim$ 10 days, in the case of Sun-like stars \citep{Owen2013,Lopez2013}. The mechanism of this photoevaporation process involves the photoionization of hydrogen from the absorption of X-rays and extreme ultraviolet radiation (XUV), which occurs at atmospheric pressures of around a nanobar, where a H/He-rich atmosphere becomes optically thin to the XUV \citep{Yelle2004,Owen2012,Murray-Clay2009}. The thermal excitation resulting from the photoionization can produce a hydrodynamic escape of the atmospheric hydrogen/helium. The impact of photoevaporation on the evolution of a short-period sub-Neptune can be substantial---low-mass planets with H/He dominated envelopes can have their envelopes completely stripped leaving only the solid core \citep{Jackson2010, Owen2012, Lopez2012}.

However, the effects that photoevaporation-driven atmospheric escape can have on these planets vary significantly based on physical parameters such as planetary radius, planet mass, host star age and incident high-energy flux. For this reason, evolution models of sub-Neptunes, which account for the effects of hydrodynamic mass-loss and thermal contraction, have been developed to study this process. Studies that have run these models over large parameter spaces in incident bolometric flux and planetary mass have indicated a threshold above which planets will not be found after $> 100$ Myr due to photoevaporation-driven mass loss \citep[e.g.,][]{Owen2013,Lopez2013,Chen2016}. The thresholds produced by these models are in good agreement with observations, which show a dearth of planets with low gravitational binding energies and large radii \citep{Lopez2016a,Owen2017,Jin2017}.

Nevertheless, the sample of planets for which masses, and in turn binding energies, are known is relatively small. This has motivated searches for regions in planet radius space with a lack of sub-Neptune sized planets in the wider \textit{Kepler} candidate dataset, where masses for most planets are unknown, but planetary radii and orbits have been measured. \citet{Sanchis-Ojeda2014} examined the population of \textit{Kepler} planets with orbital periods less than one day, and noted a clear lack of planets larger than 2 $R_{\oplus}$, suggesting that photoevaporation could be responsible for this observation. More recently, \citet{Mazeh2016} defined the shape of the desert in the period-mass and period-radius planes, and \citet{Lundkvist2016} used a sample of planets around host stars with asteroseismic observations to suggest an absence of planets with bolometric fluxes greater than 650 times the solar constant (650 $F_{\oplus}$) and radii between 2.2 -- 3.8 $R_{\oplus}$.

These studies have indicated the presence of a ``photoevaporation desert'' in the \textit{Kepler} data, or a dearth of sub-Neptunes on close-in orbits. Nevertheless, there remains much to be gleaned from observational constraints on the desert using the \textit{Kepler} data. In particular, one expects the photoevaporation desert to depend on the X-ray and XUV flux received by a planet integrated over its lifetime, which is the main driver of atmospheric escape for these planets. Studies to date have defined the desert in the parameter spaces of orbital period and present-day bolometric flux. Here, we use the X-ray observations of \citet{Jackson2012} and \citet{Shkolnik2014} to generate a lookup table of a star's lifetime-integrated X-ray luminosity as a function of its mass and age. Scaling the above by the semi-major axis of each planet of a subset of the \textit{Kepler} data selected to minimize the effects of pipeline incompleteness, we define the photoevaporation desert in a lifetime-integrated X-ray flux parameter space. This allows us to quantify clear differences between the populations of rocky and gaseous planets, as well as to define the desert in a parameter space physically tied to the photoevaporation process, facillitating comparisons with the results of photoevaporation models. Furthermore, the effect of the stellar type of a planet's host star on its propensity for photoevaporation, and on the shape of the desert in the \textit{Kepler} data has remained relatively unexplored. \citealt{Owen2013} briefly discussed these effects in a comparison of their photoevaporation model with the \textit{Kepler} data. Our analysis significantly expands on this by defining desert bounds independent of a particular photoevaporation model using statistical criteria, in addition to utilizing significantly expanded and improved datasets.

The recent significant increase in the number of known exoplanets, specifically the 1284 new confirmed planets added to the general \textit{Kepler} data by the uniform false positive probability analysis of \citet{Morton2016}, along with the factor of three improvements in the precision of planetary radii, order of magnitude improvements in stellar fluxes, and first comprehensive dataset of planet ages provided by the California-Kepler Survey \citep{Johnson2017} make this a well-timed occasion to improve observational constraints on the photoevaporation desert. With a sample size of 2341 detected planets, it is also possible to see how the desert varies as a function of the host star's spectral type.

We also investigate the photoevaporation valley, a gap in the \textit{Kepler} planet data at 1.5 -- 2 $R_{\oplus}$ first reported by \citet{Fulton2017}, in the lifetime-integrated X-ray flux parameter space, as well as its dependence on stellar type. We find evidence suggestive of the sub-Neptune valley around M \& K dwarfs, whereas previous detections have focused on Sun-like stars. The reader is referred to Fig. 10 of \citealt{Fulton2017} for an illustrated comparison of the locations of the photoevaporation desert and valley, which are two distinct phenomena.

\section{Methods}
 \label{sec:results}
\subsection{Lifetime-integrated X-ray flux as a function of stellar mass and age} 
 \label{sub:xray_lookup}
\subsubsection{Availability of extreme ultraviolet data}
With the exception of the Sun, extreme ultraviolet (EUV, 10 -- 124 nm) observations of stars are scarce. While studies have used models derived from solar observations to estimate XUV fluxes for old, Sun-like stars \citep{LecavelierDesEtangs2007, Sanz-Forcada2011}, such extrapolations are not possible for other stellar types. Thus, we focus specifically on the evolution of the X-rays. X-rays are the primary driver for hydrodynamic escape \citep{Owen2012}. Furthermore, the XUV emission for Sun-type stars \citep{Sanz-Forcada2011}, and the trends of the near and far ultraviolet for M-dwarfs \citep{Shkolnik2014}, are all suggestive of the XUV evolution qualitatively following the same saturation and decay behavior as the X-rays.

\subsubsection{Combining existing -X-ray data}
 \label{subsub:xray_data}
To track the evolution of X-ray flux as a function of time for different stellar spectral types, we use the observationally derived relations in \citet{Jackson2012} and \citet{Shkolnik2014}. \citet{Jackson2012} used X-ray survey observations of open clusters to report the ratio of the X-ray to bolometric luminosity across stellar ages of 10 Myr to 4.5 Gyr, with bins spanning stellar spectral types from K5 to F0. \citealt{Shkolnik2014} used \textit{ROSAT} (R\"{o}ntgensatellit) observations to determine the X-ray to J-band (centered at 1.25 $\mu m$) flux  ratio for M4 to K4 spectral types with stellar ages of 10 Myr to 5 Gyr. Combining these two data sets, the rough range of stellar masses for which the lifetime X-ray evolution has been measured is 0.3 -- 1.4 $M_{\astrosun}$. These data measure the steady state X-ray emission and do not account for additional X-ray emissions from large flares or coronal mass ejections.

\begin{figure}
\includegraphics[width=\linewidth]{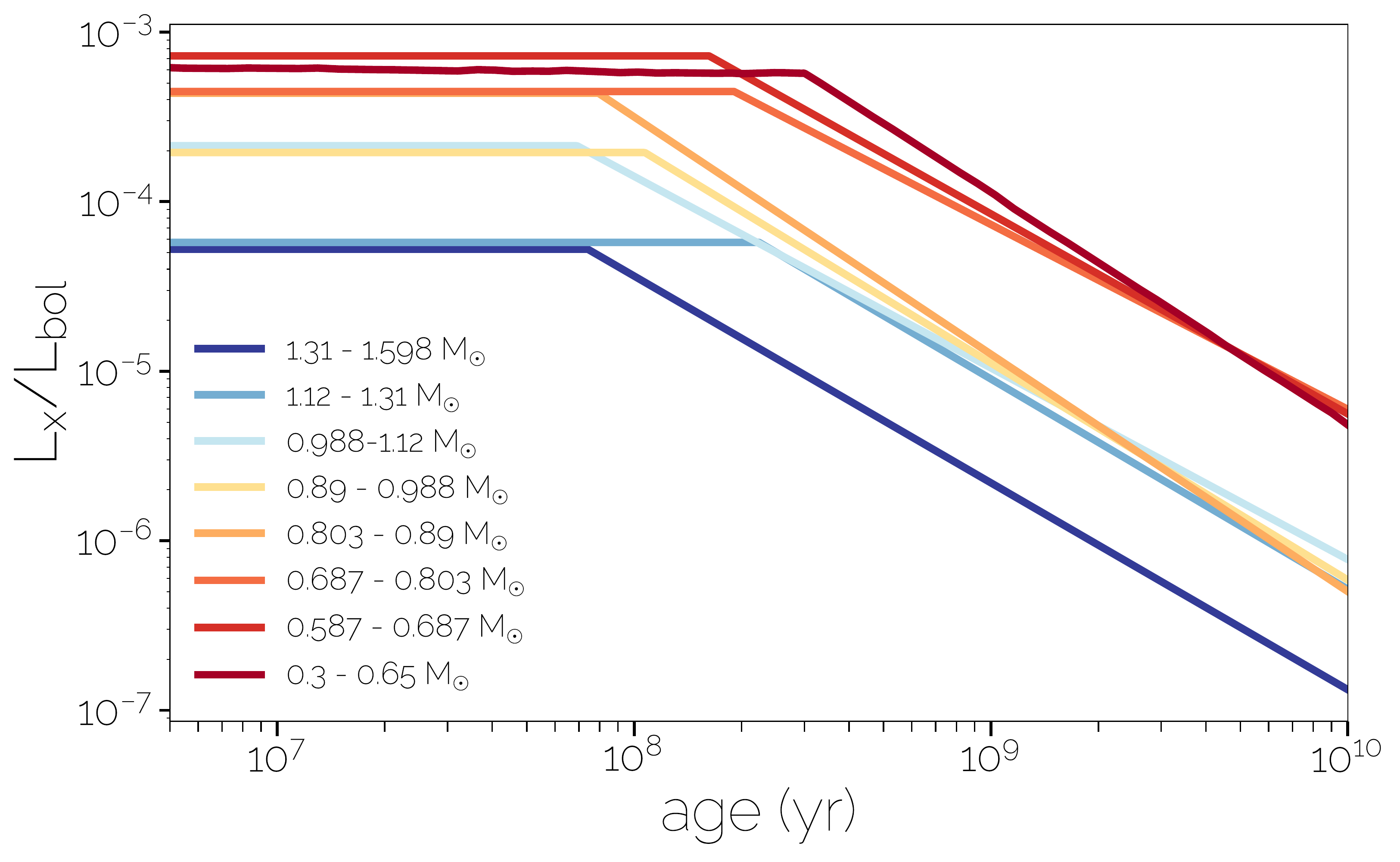}
\caption{X-ray evolution as a function of stellar type. Data are from \citet{Jackson2012} and \citet{Shkolnik2014}.}
\label{fig1}
\end{figure}

The same trend of a saturated phase of X-ray emission early in the lifetime of the star, tens to hundreds of millions years, followed by a decay is found across all stellar spectral types \textbf{(Fig. 1)}. The main difference between spectral types is twofold: as stellar mass decreases, stars saturate with a greater proportion of their luminosity in the X-rays, while the saturation phase also increases in duration \citep{Jackson2012,Shkolnik2014}. Both trends are visible in \textbf{Fig. 1}. Although this behavior is not monotonic across spectral types as a result of the uncertainties in the observations, the overall trend is apparent. We note that the binning by spectral type of \citet{Jackson2012} was carried using $B - V$ color---we convert the main sequence $B - V$ color bins to stellar mass bins solely for reference in \textbf{Fig. 1}. The conversion from $B - V$ color to stellar mass varies as a function of stellar age, and the full treatment for our calculation of lifetime-integrated X-ray fluxes is more complex and described in detail in \ref{subsub:stellar_evo}.

Both \citet{Jackson2012} and \citet{Shkolnik2014} parametrize the X-ray to bolometric luminosity ratio ($L_X/L_{bol}$) and X-ray to J-band luminosity ($L_X/L_J$) respectively, as a function of stellar age (for a given spectral type) by fitting a broken power law. The saturated phase is fit by a flat line, and the subsequent decay is fit with a power law. The uncertainties in the X-ray measurements of open clusters and field stars, which are used to define individual data points, can span orders of magnitude for the least precise measurements, with the precision varying as a function of stellar age and spectral type, both parameters that are independent variables in our analysis. In order to accurately account for these errors in our calculations, we carry out a Markov-chain Monte Carlo (hereafter MCMC) fit to all data points in each of the stellar bins defined by \citet{Jackson2012} and \citet{Shkolnik2014}. In carrying out the MCMC, we take the least squares fit of a broken power law (using a piecewise function to fit to all data points simultaneously) to the data and their errors in $L_X/L_{bol}$ to be the maximum likelihood result.

The fit itself is carried out in the log (age[yr]) and log($L_X/L_{bol}$) parameter space, such that we are fitting to the data with a piecewise function of two lines (as in \textbf{Fig. 1}), one with a slope of zero representing the saturation phase, and a line of negative slope representing the decay in X-ray luminosity as a function of time. The parameters that we fit for are the log($L_X/L_{bol}$) intercept for the line of negative slope, the logarithm of the time at which the saturation phase ends (log $\tau_{sat}$), and the slope of the line representing the X-ray luminosity decay---under the condition that the piecewise function is continuous. These fits, which are carried out in logarithmic space, are then converted to the following: the ratio of X-ray to bolometric, or J-band for \citet{Shkolnik2014}, luminosity during the saturation phase ($L_X/L_{bol}$), the time at which the saturation phase ends $\tau_{sat}$ and the power-law index $\alpha$ for the decay in $L_X/L_{bol}$ after $\tau_{sat}$. This approach returned fit parameters in many cases identical to those reported in \citet{Jackson2012} and \citet{Shkolnik2014}. So as to ensure agreement of our maximum likelihood fits with those reported in \citealt{Jackson2012}, we carried out the ``lower weighting'' of data points referenced with arrows described in \textit{Fig. 2} of \citet{Jackson2012}---we found that multiplying the errors for these data points by a factor of five produced fit parameters in closest agreement with theirs.

With our maximum likelihood fit parameters in hand, we sampled the parameter space with a Markov-chain Monte Carlo of 100 samplers over 1200 steps using the \texttt{emcee} routine of \citealt{Foreman-Mackey2012}. Discarding the first 200 steps during which the samplers have not yet sampled the full parameter space, we are left with 100,000 fits, comprising three posterior probability distributions, one for each fit parameter. We randomly sample the fit parameters for 1000 of the fits, taking note that this number of subsamples still accurately represents the shape of the original posterior probability distributions. These 1000 fits are used to account for the effects that the errors in the X-ray measurements have on our calculation of lifetime-integrated X-ray luminosities and fluxes.

\subsubsection{Stellar evolution and calculation of a lifetime-integrated X-ray luminosity lookup table}
 \label{subsub:stellar_evo}
We ultimately seek to calculate estimates of the lifetime-integrated X-ray flux for all planets in the \textit{Kepler} sample. However, the calculation of lifetime-integrated X-ray flux for each planet, as well as a Monte Carlo sampling of each planet's errors, is precluded by the time-intensive nature of the calculation. Because the resolution of the stellar masses in the isochrones that we use as inputs, as well as the ages at which lifetime-integrated X-ray fluxes have been measured, are known, we are aware of the maximum resolution at which lifetime-integrated X-ray luminosities can be accurately calculated as a function of stellar mass and age. For this reason, we precalculate a lookup table---an array of lifetime-integrated X-ray luminosities for given stellar masses and ages, which can be quickly queried for use in estimating the lifetime-integrated X-ray flux for each \textit{Kepler} planet.

As discussed in section \ref{subsub:xray_data}, the data on X-ray evolution are reported in terms of $L_X/L_{bol}$ or $L_X/L_J$. As such, it becomes necessary to model the temporal evolution of the bolometric and J-band luminosities for the stellar types of interest, in order to determine the temporal evolution of the X-ray luminosity itself. Stellar isochrone models are well suited for this, and we use the \textit{PARSEC} isochrones obtained from the \texttt{CMD 2.8} web interface \citep{Bressan2012,Chen2014}. We utilize the \textit{PARSEC} isochrones for their fine mass resolution ($\sim 0.03$) over our $0.3 < M_{\astrosun} < 1.4$ range of interest, as well as the availability of tracks for stars with non-solar metallicities.

In generating our lookup table, we define 40 linearly spaced mass bins over the solar mass range $0.3 < M_{\astrosun} < 1.4$, as well as 50 different stellar ages which we integrate up to. 20 ages are logarithmically spaced between 200 Myr (limited on the low end by the lowest ages in the \textit{PARSEC} isochrones) and 1 Gyr, and the final 30 are linearly spaced fom 1 -- 10 Gyr so as to finely sample the age range which most \textit{Kepler} planets fall in (\citet{Johnson2017}, also see \textbf{Fig. 3}). Finally, 5 bins in metallicity are specified at $Z =$ 0.0025, 0.0075, 0.0125, 0.0175 and 0.03.

For each mass, age and metallicity bin, we track the star's luminosity and $B - V$ color at each time-step in the isochrone (spaced as log(time[yr]) = 0.03). At each time-step, we use the $B - V$ color to place the star in the appropriate bin of X-ray evolution as described in section \ref{subsub:xray_data}. We then multiply the isochrone luminosity by one of the $L_X/L_{bol}$ ratios among the 1000 fits generated for this bin in section \ref{subsub:xray_data}. This leaves us with $L_X$ as a function of time, which we integrate over the lifetime of the star by quadrature (a numerical integration using polynomial interpolation functions) to get a lifetime-integrated X-ray luminosity. For stars of $M_* > 0.9 M_{\astrosun}$ that are older than 1 Gyr, we monitor the evolution of the star's stellar effective temperature to determine whether it leaves the main sequence during its lifetime. We consider a greater than 10\% change in stellar effective temperature vs. log(time[yr]) to be a signature that the star has left the main sequence. This criteria accurately captures the point in time before dramatic changes to the star's other physical parameters, e.g. $R_*$, also occur. We stop the integration of lifetime-integrated X-ray flux at the time that the star has left the main sequence, due to the lack of comprehensive observations of X-ray emissions from post main-sequence stars. This process of calculating lifetime-integrated X-ray luminosity is repeated for all mass, age and metallicity bins, and then repeated for each one of the 1000 sets of X-ray evolution fit parameters from section \ref{subsub:xray_data}. This effectively provides 1000 realizations of the lookup table that sample the extent of the errors in the X-ray measurements.

\begin{figure*}
\includegraphics[width = \textwidth]{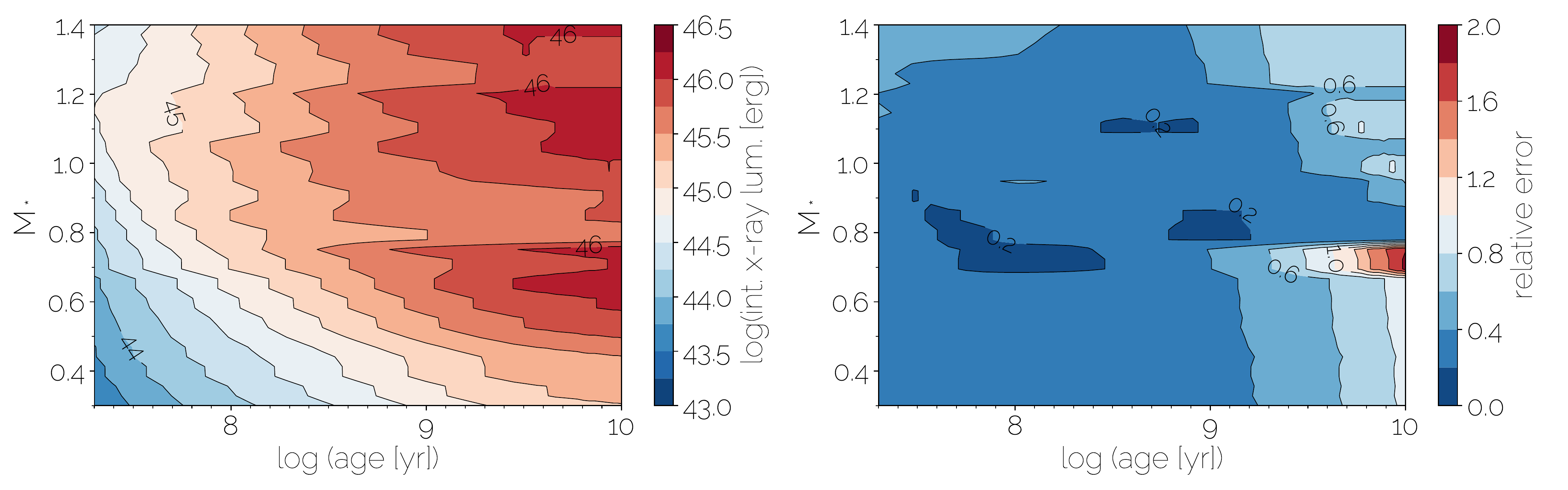}
\includegraphics[width = 0.5\textwidth]{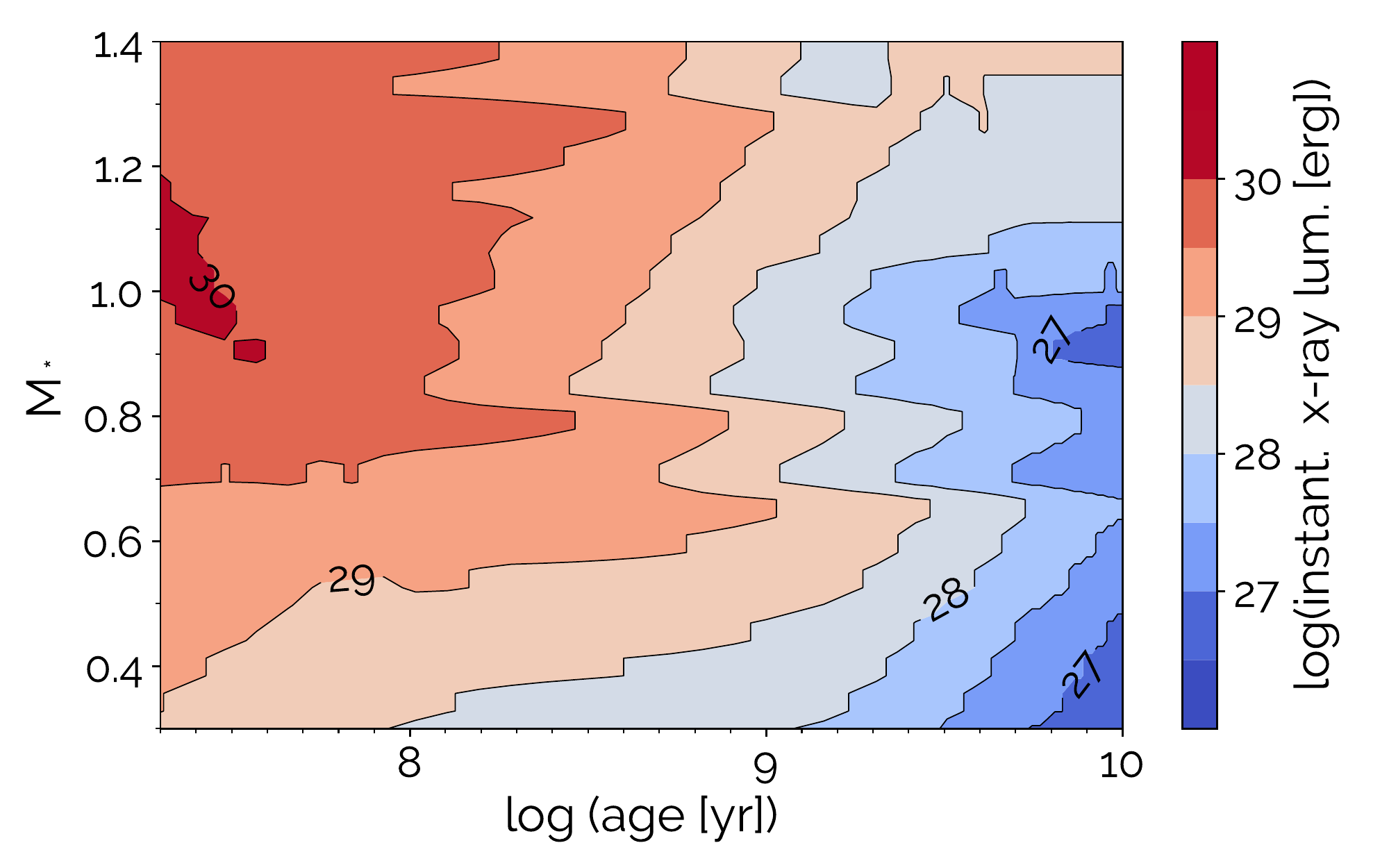}
\caption{a) Lookup table of integrated X-ray luminosity as a function of
stellar mass and age, using fits to the observational relations of \citet{Jackson2012} and \citet{Shkolnik2014}. This table is for a stellar metallicity of $Z =$ 0.0175, and shows the mean luminosities of 1000 realizations of fits to the data. b) Relative errors on the X-ray luminosity calculated from the 1000 realizations. c) Lookup table of instantaneous X-ray luminosity using the same methodology as for \textbf{Fig. 2a}, but without the time integration. \textit{These lookup tables are available on the website of Laura Kreidberg.}}
\label{fig2}
\end{figure*}

In \textbf{Fig. 2}, we show our lookup table for our closest-to-solar metallicity bin ($Z =$ 0.0175). \textbf{Fig. 2a} shows the mean lifetime integrated X-ray luminosity of the 1000 realizations in X-ray evolution fit parameters, while \textbf{Fig. 2b} shows the relative error in the lifetime-integrated X-ray luminosities. The physical implications of the results in the lookup table will be discussed in section \ref{subsub:xray_vs_mass}. Over the majority of the parameter space, relative errors are around 30\%, although for stars of $M_* \sim 0.7 M_{\astrosun}$ older than 3 Gyr, relative errors are greater than 100\%. This is due to poor constraints on the post-saturation phase X-ray evolution for this particular bin of $0.935 < B - V < 1.275$ (see \citet{Jackson2012} \textit{Fig. 2f}).

For reference, and to facilitate comparisons with the results of evolution models, we have also show a lookup table of instantaneous X-ray luminosity (\textbf{Fig. 2c}). This uses the same methodology used to generate the lifetime integrated X-ray luminosity lookuptable, but without an integration over time.

Although we utilize and report results using the \textit{PARSEC} isochrones due to their mass and metallicity resolution, as well as their direct outputting of $B - V$ color, we have verified our results by repeating the above analysis using the \citet{Baraffe2015} stellar isochrones, combined with estimates of $B - V$ color from model stellar effective temperature using the observations of \citet{Pecaut2013}. We find that our results are largely independent of the isochrones used.

\begin{figure}
\includegraphics[width = 0.5\textwidth]{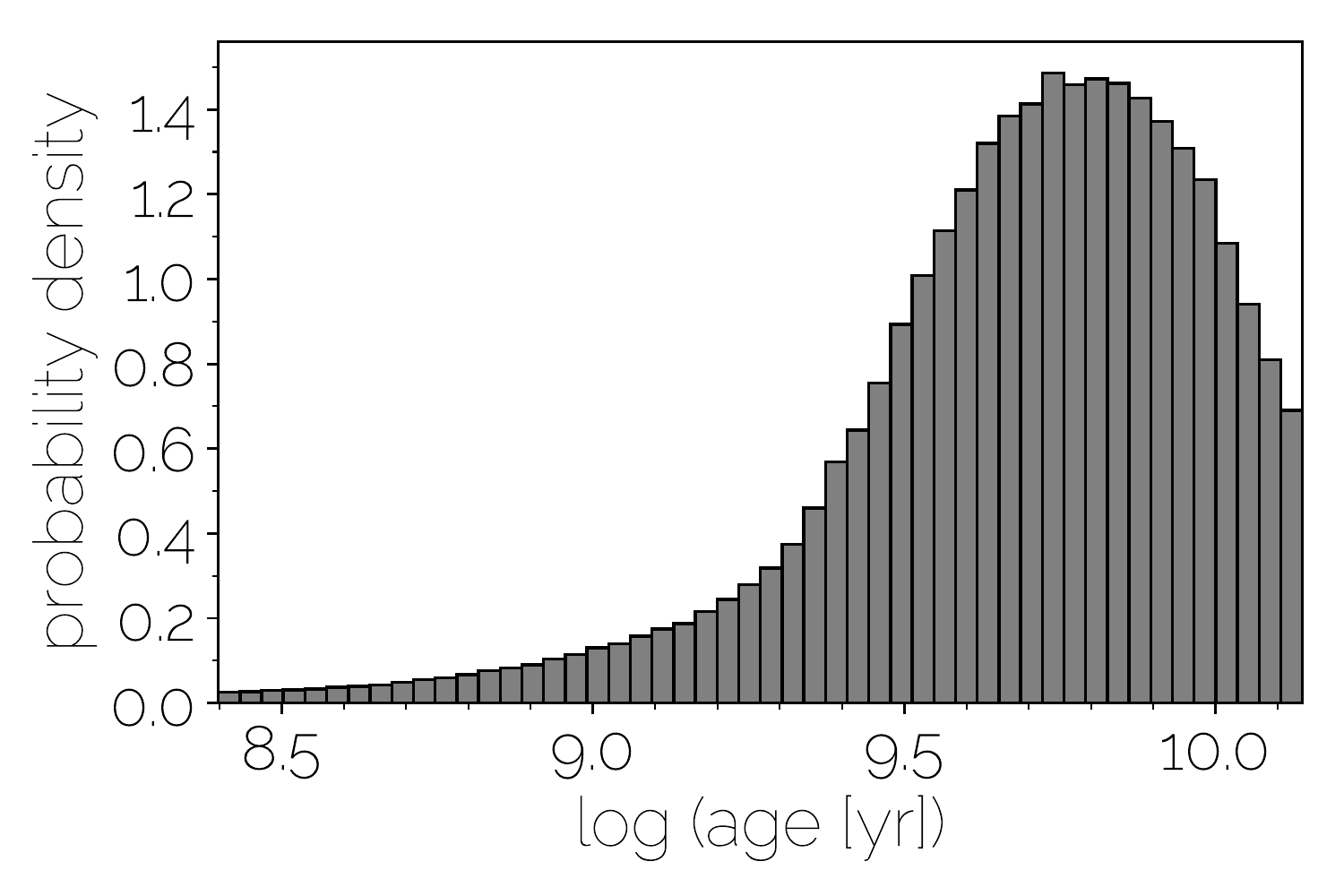}
\caption{A histogram of the relative frequencies of all host star ages from the California Kepler Survey (CKS, citet{Petigura2017,Johnson2017}). In order to account for the errors in the determination of ages, 1000 realizations of the age of each planet are included in the histogram.}
\label{fig3}
\end{figure}

\subsection{Lifetime-integrated X-ray fluxes for \textit{Kepler} candidates}
 \label{sub:apply_kepler}
\subsubsection{Kepler data}
 \label{subsub:kepler_data}

In this section, we calculate the lifetime-integrated X-ray flux for stars in the \textit{Kepler} sample. We utilize two sets of \textit{Kepler} data. For the purposes of defining the photoevaporation desert with the greatest possible precision, we use data from the California Kepler Survey \citep{Petigura2017,Johnson2017}. The California Kepler Survey (abbreviated CKS) used follow-up spectroscopic observations to improve the precision on the stellar parameters of 1305 \textit{Kepler} stars, and in turn the measurements of 2025 \textit{Kepler} planets. The CKS measurements have resulted in a factor of $\sim$3 improvement in the precision on measurement of stellar masses, as well as a factor of $\sim$2 improvement in stellar radius precision that has resulted in a factor of $\sim$2 improvement in precision for the planetary radii (using the planet-to-star radius ratios from the latest Q1-17 DR 25 \textit{Kepler} catalog, \citet{Johnson2017}). Furthermore, the CKS is the first comprehensive survey of \textit{Kepler} stellar ages (see \textbf{Fig. 3}), increasing the sample size of stars with ages by almost two orders of magnitude over previous studies. This enables, for the first time, calculating estimates of parameters such as lifetime-integrated X-ray flux for a majority of planets in the \textit{Kepler} dataset.

Because the CKS is a magnitude limited survey (less than magnitude 14.2 in the \textit{Kepler} bandpass, \citet{Petigura2017}), it excludes most low-mass stars from the \textit{Kepler} sample, and consists largely of Sun-like stars. To explore differences in the photoevaporation desert and valley as a function of stellar type, we use the latest release of the general \textit{Kepler} dataset: the Q1-17 DR 25 catalog \citep{Thompson2018}. The recent calculation of uniform false positive probabilities (FPP) for all \textit{Kepler} Objects of Interest in the Q1-17 DR 25 catalog using the methodology of \citealt{Morton2016} has resulted in a large increase in the number of confirmed \textit{Kepler} planets, totaling 3182 planets. The stellar parameters utilize those corrected from the original DR 25 catalog, i.e. the DR 25 Supplemental
Stellar data.

\subsubsection{Propagating the \textit{Kepler} planet uncertainties through our calculation of lifetime X-ray fluxes}
\label{subsub:kepler_mc}

The relative errors on the properties of the \textit{Kepler} planets and host stars can be large. Even with the improved precision of the California Kepler Survey, the typical 5\% errors on stellar masses, 11\% uncertainty factor on stellar radius, and factor of two uncertainty in ages can result in significant variations in estimates of lifetime-integrated X-ray flux for a given planet. The effect of these errors on defining a region free of planets (the photoevaporation desert) must be taken into account, especially when dealing with samples limited to several dozen to a couple hundred planets (the reason for the smaller samples of planets will be discussed in section \ref{subsub:complete}). The consideration of errors becomes even more important when using the regular \textit{Kepler} sample to compare properties dependent on host star stellar type, in which typical errors on planet radii are $\sim 40\%$ \citep{Johnson2017}.

In consideration of these errors, as well as the fact that our calculations of lifetime-integrated X-ray luminosities rely on numerical integration of binned observational data (making it impractical to propagate errors), we take a Monte Carlo approach to modeling the errors on planets. We replace each planet in the specific dataset being considered (CKS or Q1-17 DR 25) with 1000 ``fractional planets'', each weighted to be 1/1000$^{th}$ of a planet. The physical parameters for each of the 1000 planet fractions and their respective host star are determined by taking posteriors directly from the relevant table: the Q1-17 DR 25 stellar catalog \citep{Mathur2016}, the Q1-17 DR 25 planet catalog \citep{Hoffman17}, or the California Kepler Survey (\citet{Petigura2017,Johnson2017}, posteriors were provided by Erik Petigura upon request). The lifetime-integrated X-ray flux for each fractional planet is determined by using the lookup table described in section \ref{subsub:stellar_evo} to get a lifetime-integrated X-ray luminosity for its Monte Carlo'd host star. 1000 realizations of the lifetime X-ray luminosity lookup table were previously generated to account for uncertainties in the X-ray observations. Each of the 1000 realizations of the lookup table are used for each of the 1000 planet fractions representing one planet, such that the uncertainties in the lifetime X-ray luminosity are propagated. The time-integrated X-ray luminosity for a fractional planet is then divided by the posterior for its semi-major axis to get a time-integrated X-ray flux.

For the CKS data, posteriors of the ages of each planet's host star are available, and drawn from for each fractional planet. In the case of the Q1-17 DR 25 catalog, ages have not been measured. While many of the G type host stars in the catalog have ages that have been measured by the CKS, the majority of the M \& K type stars, which we wish to compare to, have no age information. In order to treat the ages, and in turn the lifetime integration over X-ray luminosity, consistently for all stellar types in the Q1-17 DR 25 catalog, we take the approach of randomly drawing an age for the host star of each fractional planet in the catalog from the full distribution of ages of the CKS sample (whereby the errors in the CKS ages are accounted for by including 1000 age posteriors for each star in the CKS sample). Because the ages measured by the CKS form a single unimodal distribution around $\sim 6$ Gyr (\textbf{Fig. 3}), a random drawing of ages for each host star in the Q1-17 DR 25 catalog should produce similar average behavior over a large sample.

\subsubsection{Selecting a subsample minimally affected by variations in completeness near the desert edge}
\label{subsub:complete}
Using the lifetime-integrated X-ray fluxes calculated for each of the 1000 fractional planets, we calculate planet occurrences in the parameter space of planet radius ($R_p$) vs lifetime-integrated X-ray flux.

Planet occurrence calculations must account for the completeness of the \textit{Kepler} dataset \citep{Howard2012}, i.e. whether every planet is likely to have been detected around the sample of stars of interest. We therefore focus our analysis on a subsample of the \textit{Kepler} dataset that is nearly complete. To obtain this sample, we follow the methodology of \citet{Wolfgang2015} and perform a series of cuts in orbital period, planet radius, stellar radius, and stellar noise (combined differential photometric precision, or CDPP). Specifically, we restrict the range of orbital periods to $<10$ days, the planet radii to $R_p > 1.2 R_\oplus$, the stellar radii to $R_* < 1.2 R_s$, and the CDPP binned to the timescale of the planet’s transit $< 200$ ppm. Our CDPP cut is less restrictive than that of \citet{Wolfgang2015} (200 ppm versus 100 ppm) because we are focusing on planets with shorter orbital periods (10 versus 25 days).

To assess the completeness of this sub-sample, we compared it to estimates of completeness from \citet{Thompson2018}.  For the subsample of planets orbiting FGK dwarfs in the Q1-17 DR 25 catalog, with period $<100$ days and $R_p < 6R_{\oplus}$ (most similar to the sample in our study), \citet{Thompson2018} found the lowest completeness was found to be 89\%, with the highest being 96\% \citet{Thompson2018}. This equated to a 7\% variation in completeness across the sample. Because sample incompleteness increases with orbital period, and the largest period in our sample is 10 days versus the 100 days in \citet{Thompson2018}, these completeness estimates are conservative.

Relative to other sources of uncertainty in our sample, the estimated errors due to pipeline incompleteness are small. The relative error due to lifetime-integrated X-ray flux is near 30\% (see \textbf{Fig. 2b}), and the relative errors due to binomial statistics are greater than 20\% (\textbf{Fig. 7b}). We therefore conclude that the effects of pipeline incompleteness will not significantly bias the main conclusions of our analysis.

The number of planets, as well as stars, left by the above cuts are tabulated in \textbf{Table 1} for the two datasets we use, as well as their stellar subsamples. Around 150 planets form the sample with host stars of all spectral types for both the CKS and the Q1-17 DR 25 datasets. The extremely small sample size of planets around F type stars (6.2) is a result of a large fraction of these stars having left the main sequence for the \textit{Kepler} ages, and in turn being discarded. We include our analyses on the F type stars for reference, but focus our interpretations on the M \& K and G type subsamples due to their larger sample sizes.

\begin{deluxetable*}{lcr}
\tablecolumns{3}
\tablewidth{0pt}
\tablecaption{The number of planets as well as the average number of stars each planet could be detected around (this varies by planet, based on its host star's noise) for each of our samples of interest. These samples are the California Kepler Survey and the Q1-17 DR 25 sample including its stellar subsamples.}
\tablehead{\colhead{Data set} & \colhead{Planet sample size} & \colhead{Average stellar sample size}}
\startdata
Data set & Planet sample size & Average stellar sample size \\
California Kepler Survey & 155.0 & 8.1 $\times 10^4$ \\
Q1-17 DR 25 & 246.8 & 6.6 $\times 10^4$ \\
\qquad M \& K subsample & 69.2 & 8.5 $\times 10^3$ \\
\qquad G subsample & 123.6 & 4.5 $\times 10^4$ \\
\qquad F subsample & 6.2 & 1.4 $\times 10^4$ \\
\enddata
\tablenotetext{*}{The number of planets can deviate from an integer because we consider individual fractional planets to sample measurement uncertainities, these are weighted such that 1000 fractional plaenets are the equivalent of one planet.}
\end{deluxetable*}

\subsection{Constraining the shape of the photoevaporation desert}

\subsubsection{Calculating occurrence}
\label{subsub:occ_calc}

We create two grids to calculate planet occurrence: one in the parameter space of planet radius and lifetime-integrated X-ray flux, and the other in planet radius and bolometric flux space, for comparison with previous studies. We define our grids so that the parameter space defined by our bounds in bolometric flux are equivalent to the bounds defined for lifetime-integrated X-ray flux, for a solar mass star of age 5 Gyr (defined at the low end to correspond to our 10 day period cut, at the high end based on the maximum lifetime-integrated X-ray flux calculated for the \textit{Kepler} planets). To calculate the occurrence within a given grid cell, we adopt the methodology of \citealt{Howard2012}. We reproduce equation (2) from \citet{Howard2012}, where the average planet occurrence within a cell of $R_p$ and lifetime-integrated X-ray flux is:

\begin{equation}
  \label{occ}
  f_{cell} = \sum\limits_{j=1}{n_{pl,cell}} \frac{1/p_j}{n_{*,j}}
\end{equation}

where $p_j = (R_*/a)_j$ is the transit probability for fractional planet $j$, with $a$ being its semi-major axis. Because we sum over fractional planets, we weight each fractional planet to be 1/1000$^{th}$ of a full planet. $n_{*,j}$ is the number of stars that pass the cuts described in section \ref{subsub:complete} for the transit duration of the specific fractional planet under consideration.

Again, following \citet{Howard2012}, we calculate the error in $f_{cell}$ using binomial statistics. The standard deviation of $f_{cell}$ is calculated from the probability distribution of drawing $n_{pl,cell}$ planets from $n_{*,eff,cell} = n_{pl,cell} / f_{cell}$ effective stars:

\begin{equation}
   \label{occ_std}
   f_{cell,std} = \sqrt{n_{*,eff,cell}f_{cell,std}(1 - f_{cell,std})}
\end{equation}

 In reporting occurrences, we choose to report occurrence densities \citep{Foreman-Mackey2014}, where we are dividing the occurrence in each grid cell by the area of the grid in $d R_p$ $d$Flux. This way, the values that we report are not dependent on the specific grid resolution that is chosen.

\subsubsection{Characterizing the occurrence desert in two dimensions}
\label{subsub:define_desert}
We wish to search for a region of the parameter space with no planets, an occurrence desert. We expect a desert to exist for gaseous planets at high fluxes due to atmospheric escape \citep[e.g.,][]{Owen2013,Lopez2013,Chen2016}. Given the large uncertainties on planet properties and fluxes, no region of the occurrence grid is exactly zero. To define the desert, we take the largest region of the occurrence grid that is consistent with zero at two sigma confidence. To find this region, we adopt the following algorithm. We start by taking note of the planet occurrence and its 2$\sigma$ deviation in the cell of largest planetary radii (3.9 -- 5 $R_{\oplus}$) and highest flux value (i.e. the upper right corner of the plots in \textbf{Fig. 7}). Due to the low occurrence and large uncertainties arising from the small number of planets in this cell (typically a handful of fractional planets), its occurrence is always consistent at the 2$\sigma$ level with an occurrence of zero. This cell is considered the starting point for our desert. We then consider the occurrences in each of the grid cells sharing one border with our initial desert grid cell. The cell with the lowest occurrence is added to the desert. The total occurrence in the new desert consisting of two cells is calculated, and the standard deviation of the desert occurrence is recalculated using binomial statistics for all fractional planets in the desert grid cells, using equation \ref{occ_std}. This process is repeated until we have drawn the largest possible region that is still consistent with an occurrence of zero at the 2$\sigma$ level. We define this region to be the desert.

We initially calculate an occurence grid with a resolution chosen such that the largest relative errors in occurrence per grid cell for the full CKS and Q1-17 DR 25 datasets are one order of magnitude; we use this grid in studies of the photoevaporation desert. In investigating the photoevaporation valley, we select a much finer resolution that lends itself to much larger relative errors in individual grid cells. However, as our interest for the valley is in occurrence summed over the radius axis, the magnitude of the relative errors for the summed occurrences still meet our criteria of $<$ 1 dex.

\subsubsection{Characterizing the occurrence desert in one dimension}
\label{subsub:cmd}
We facilitate comparison of the photoevaporation desert among different parameters, present-day bolometric flux vs lifetime-integrated X-ray flux, as well as among different subsamples of stellar type and planet radius, by examining our full data set of planets in one dimension. We take all fractional planets from our data set of interest with $R_p$: $1.8 < R_{\oplus} < 4$ (our definition of sub-Neptunes), and generate a cumulative distribution function (CDF) over the flux dimension, beginning at the fractional planets with highest fluxes.

We wish to compare the CDFs for bolometric flux vs lifetime-integrated X-ray flux, which are parameters with different units. For this comparison, the bolometric flux data are renormalized to the lifetime-integrated X-ray flux data such that they have identical values at the 10th and 90th percentile of flux.

To determine whether the given CDFs in different physical parameter spaces, or for different subsamples, have statistically significant differences, we utilize the Anderson-Darling test (hereafter AD test) for two samples. The AD test was developed as an alternative to the Kolmogorov-Smirnov test. The AD test was designed to be sensitive to differences in the tails of two CDFs, making this test appropriate for our analysis of the photoevaporation desert, which is represented in the high flux tail. The Kolmogorov-Smirnov test is most sensitive to differences in the center of the distribution \citep{Anderson1954}. Nevertheless, we confirm our results regarding statistical differences in the distributions with the Kolmogorov-Smirnov test and find consistent conclusions. We choose, however, to report results for the AD test, for the reasons cited above.

To compare the shape of the transition from the desert to the center of the planet distribution, we also consider the probability distribution function (PDF)---simply histograms of the planet distributions in flux space.

\begin{figure}
\includegraphics[width = 0.5\textwidth]{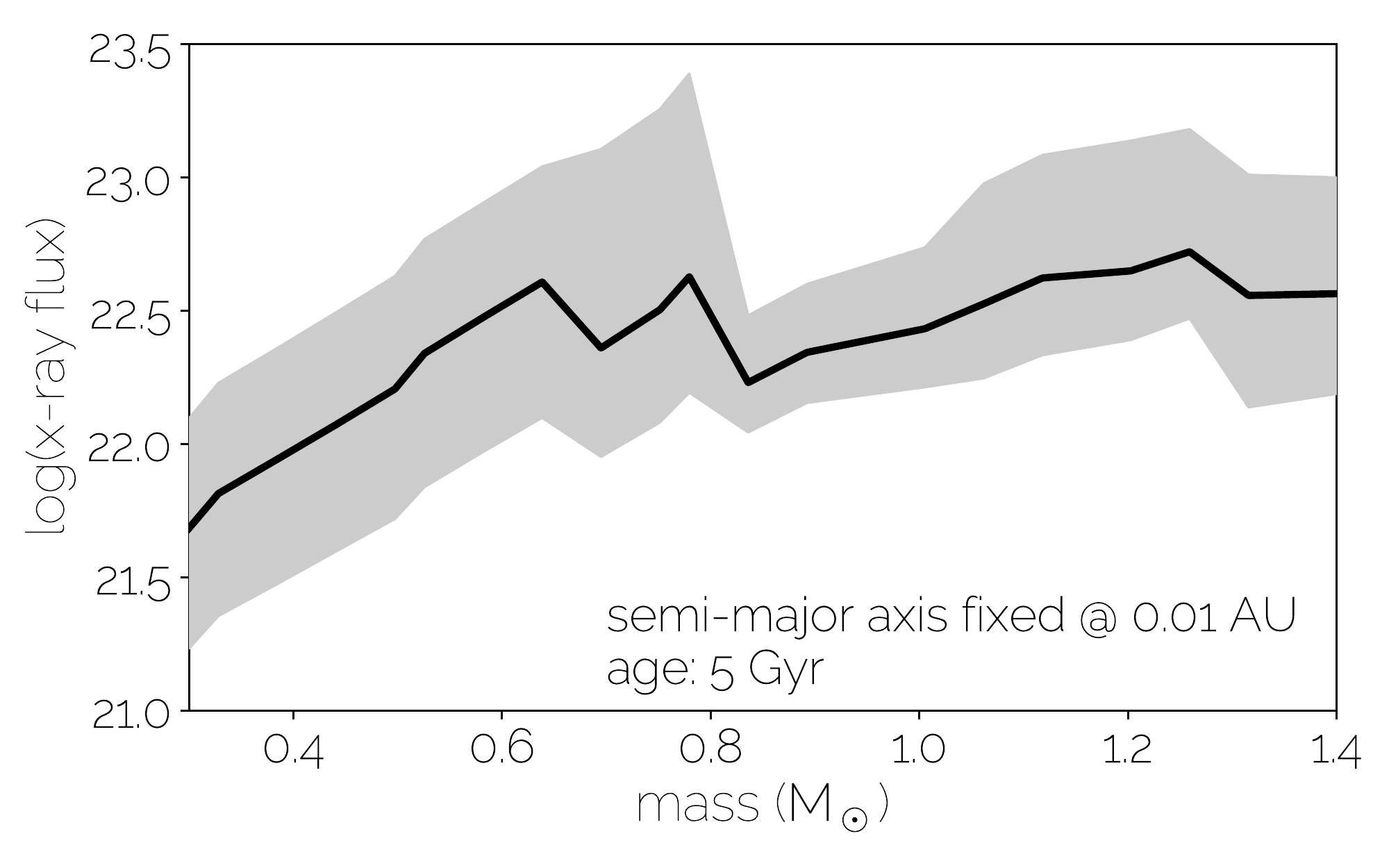}
\caption{Integrated X-ray flux over a lifetime of 5 Gyr for a planet at 0.01 AU, as a function of stellar type. The shaded error bars denote 2$\sigma$ uncertainties calculated from a Monte Carlo sampling of the \citet{Jackson2012} and \citet{Shkolnik2014} data.}
\label{fig4}
\end{figure}

\begin{figure}
\includegraphics[width = 0.5\textwidth]{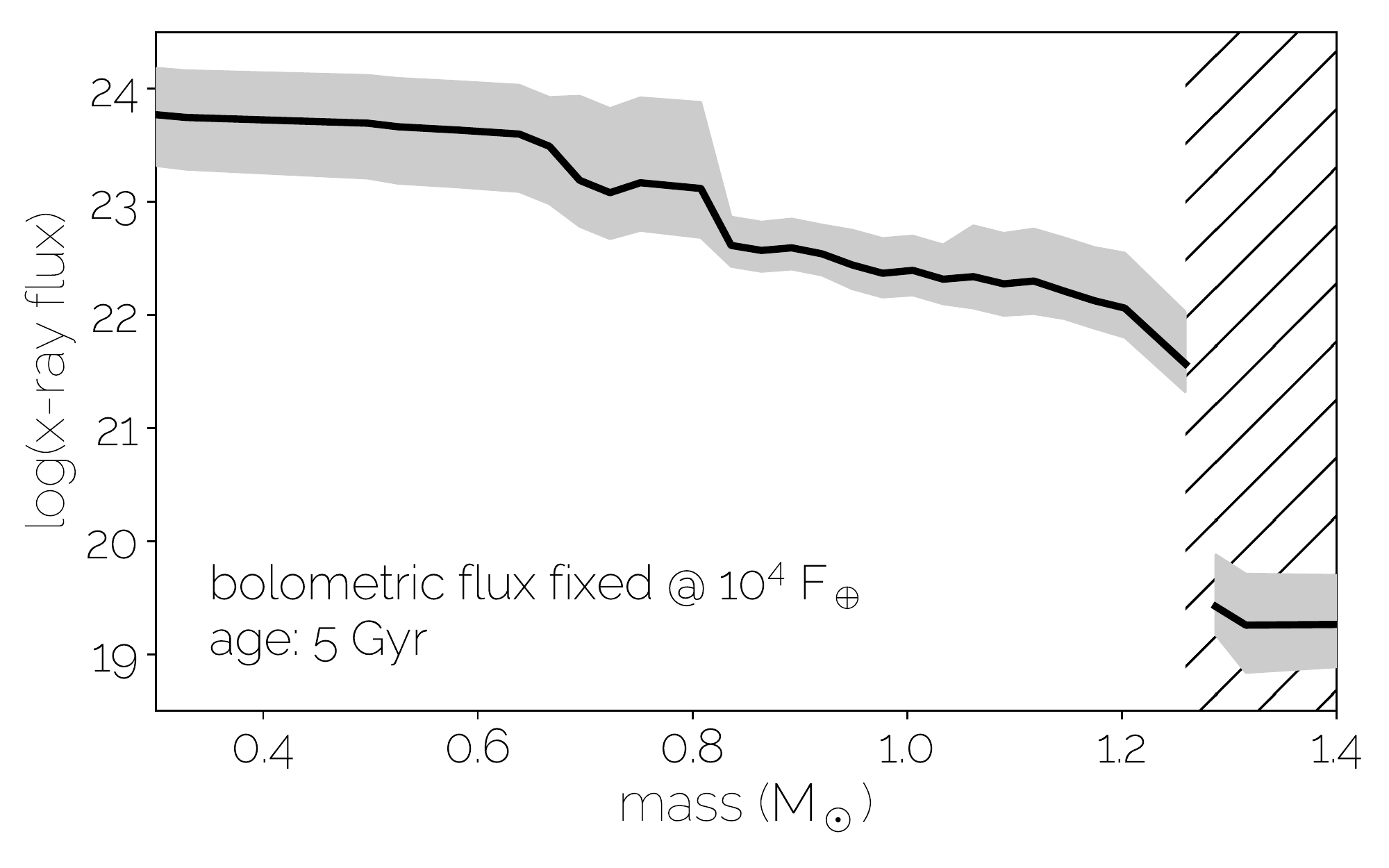}
\caption{Integrated X-ray flux over a lifetime of 5 Gyr, for a planet with its 5 Gyr bolometric flux fixed at $10^4 F_\oplus$, as a function of stellar type. Note that the semi-major axis at which the planet is placed must be varied depending on the stellar type of the host star. The shaded error bars are calculated in the same manner as in \textbf{Fig. 2} and denote denote 2$\sigma$ uncertainties. Stars within the hatched region have already left the main-sequence at an age of 5 Gyr.}
\label{fig5}
\end{figure}

\section{Results}
 \label{results}
\subsection{Lifetime-integrated X-ray flux as a function of stellar mass}
 \label{subsub:xray_vs_mass}

We examine the dependence of lifetime-integrated X-ray luminosity on stellar mass and age, to see in detail how host star properties can affect the propensity for a planet to photoevaporate. In \textbf{Fig. 4}, we use the results from the lookup table discussed in section \ref{subsub:stellar_evo} to examine how the lifetime-integrated X-ray flux would vary as a function of stellar mass for a planet of age 5 Gyr fixed at a semi-major axis of 0.01 astronomical units (AU). The shaded error bars denote the 2$\sigma$ error in the lifetime-integrated X-ray fluxes, as determined by our Markov Chain Monte Carlo fits to the X-ray data.

The lifetime-integrated X-ray flux increases as the stellar mass increases, which is not surprising given that the bolometric luminosity increases with mass. What is not quite as intuitive, is that the lifetime-integrated X-ray flux changes by only an order of magnitude for stars with 0.3 $< M_\odot <$ 1.4, despite the fact that the bolometric luminosity varies by two orders of magnitude over this mass range. Almost all of the drop in lifetime X-ray luminosity with lower stellar mass happens for the M-dwarfs $<$ 0.6 $M_{\astrosun}$. Above this mass, the lifetime X-ray luminosity is roughly flat with stellar mass since increases in the saturation level and lifetime (see \textbf{Fig. 1}) roughly balance out decreases in bolometric luminosity. Meanwhile below that mass, lifetime X-ray luminosities not only decrease as a function of mass, but also become strongly age-dependent even at several Gyr, likely due to their longer activity lifetimes combined with their continued bolometric luminosity evolution from Kelvin Helmholtz contraction (\textbf{Fig. 4}).

\textbf{Fig. 5} further emphasizes the implications of the proportionally high X-ray luminosities of low-mass stars. In this plot we have fixed the present-day bolometric flux at $10^4$ $F_{\oplus}$ for a planet of age 5 Gyr. In order to fix the bolometric flux, the semi-major axis of the planet must be varied for different stellar types. For a planet of age 5 Gyr, the lifetime-integrated X-ray flux is two orders of magnitude greater if it orbits a 0.3 $M_{\astrosun}$ star when compared to a star of mass 1.2 $M_{\astrosun}$. This plot provides an intuition for how a plot of the \textit{Kepler} planets in radius vs. bolometric parameter space (e.g Fig. 2 of \citet{Lundkvist2016}) will be affected when the independent variable is changed to lifetime-integrated X-ray flux. For a plot calibrated to have the same parameter space in bolometric flux and lifetime-integrated X-ray flux (as we have done) for a solar mass star, all planets orbiting low-mass stars will move to higher lifetime-integrated X-ray fluxes relative to those emitted by a Sun-like star, while all planets orbiting high-mass stars will move to to lower lifetime-integrated X-ray fluxes. We predict that this shuffling of planets will change the shape of the planet distribution, as well as the shape of the photoevaporation desert.

The dramatic drop off in integrated X-ray flux for stellar masses greater than $M_{\astrosun} = 1.2$ in \textbf{Fig. 5} is the result of these stars leaving the main sequence before 5 Gyr. When the stars leave the main sequence, their luminosity increases dramatically, necessitating the placement of the planet at comparatively long semi-major axes for the calculation of a bolometric flux at 5 Gyr of $10^4$ $F_{\oplus}$. Combined with the fact that we stop time-integration of the host star's X-ray luminosity when it leaves the main sequence, the resultant time-integrated X-ray fluxes are lower by $\sim2$ orders of magnitude compared to those stars that are still on the main sequence.

\begin{figure*}
\gridline{\fig{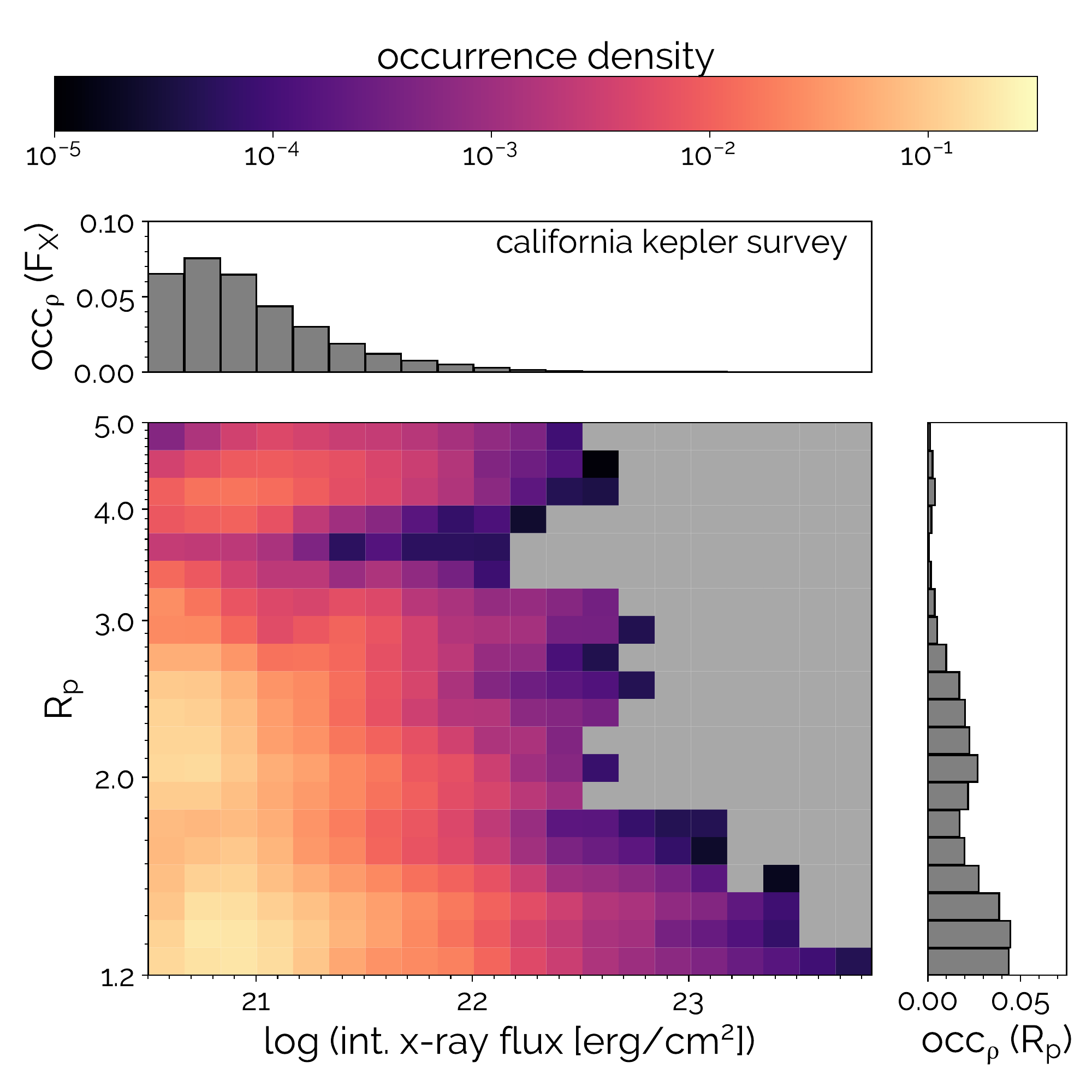}{0.5\textwidth}{a)}
            \fig{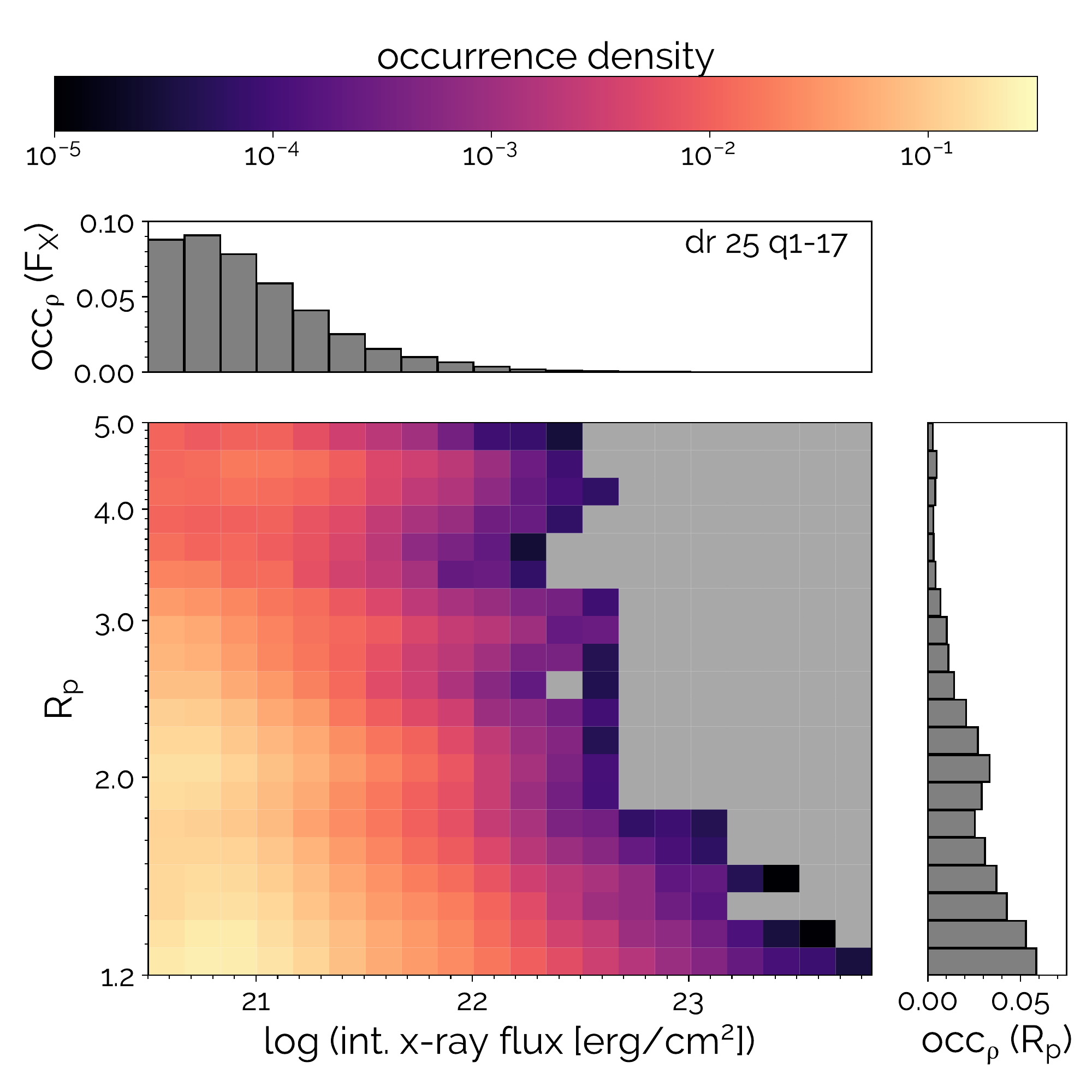}{0.5\textwidth}{b}}
\caption{a) Occurrence densities for the subsample of the California Kepler Survey. Summed occurrences over planet radius and integrated X-ray flux are shown in the 1D histograms along the plot edges, and use the same axes as the two dimensional occurrence plot. The grey shaded regions denote grid cells in which there are no data. b) The same plot, but for the subsample of the Q1-17 DR 25 data.
confirmed planets.}
\label{fig6}
\end{figure*}

\subsection{Comparison of the California Kepler Survey and Q1-17 DR 25 dataset}
 \label{sub:data_comp}
The CKS sample has more precise stellar parameters, whereas the sample size of confirmed planets in the Q1-17 DR 25 catalog is larger. We use the CKS sample to evaluate the shape of the occurrence desert at higher precision for Sun-like stars, and use the Kepler Q1-17 DR 25 sample to explore the desert as a function of stellar spectral type. In \textbf{Fig. 6}, we compare the calculated occurrences for the subsample of these datasets that we have chosen to minimize possible effects from incompleteness when compared with other sources of error (see section \ref{subsub:complete}). The purpose of this plot is to make note of systematic differences in the dataset that may affect our analyses. We note that in two dimensions, we consider the grid resolution of this plot to be oversampled in that many individual cells will have relative errors in occurrence greater than 10. We focus our attention on the occurrences summed over radius and lifetime-integrated X-ray flux (shown as histograms on the edges of the plot, with the same axes as the 2D plot). The bound of the plot on the low flux end has been set for the lifetime-integrated X-ray flux for a planet on a 10 day period, around a solar mass star with age of 5 Gyr, as a result of our cut to minimize pipeline incompleteness effects that eliminates all planets on greater than 10 day orbits. Any drop off in occurrence on the low flux end over the parameter space of interest is very likely the result of our period cut and we consider this to be nonphysical.

 No systematic differences are apparent. In both datasets, the peak occurrence as a function of lifetime-integrated X-ray flux occurs at $\sim 10^{20.7}$ erg/cm$^2$. As a function of radius, the peak occurrence in the CKS data occurs at 1.3 $R_{\oplus}$, while the Q1-17 DR 25 catalog peaks at 1.2 $R_{\oplus}$. Although reporting the occurrence density values in two dimensions is avoided due to our grid oversampling the parameter space, perhaps the most noticeable difference between the two datasets is visible just by considering relative variations in occurrence. The Q1-17 DR 25 sample exhibits more smearing of occurrence densities across both dimensions, while subtle variations that separate different planet populations are more important in the CKS data. This is a result of the factor of $\sim$2 -- 3 greater uncertainties in the Q1-17 DR 25 dataset. These uncertainties were discussed in section \ref{subsub:kepler_data}, and those with respect to $R_p$, and $M_*$ are of particular consequence, with the latter having a large affect on the calculation of the lifetime-integrated X-ray flux. The splitting of the planet population by a ``photoevaporation valley'', or a local minimum of planet occurrence at $R_p = 1.7 R_{\oplus}$ into two populations, one with peak occurrence at $R_p = 2.3 R_{\oplus}$, and another with peak occurrence at $R_p = 1.3 R_{\oplus}$ and at a higher value in the flux domain is consistent with the results of \citet{Fulton2017} (see their \textit{Fig. 10}). Nevertheless, the valley is still apparent in the Q1-17 DR 25 data set. The fact that it is still observable allows us to comment on its variation as a function of stellar type, which we will discuss in section \ref{sub:res_valley}.

\begin{figure*}
\includegraphics[width = \textwidth]{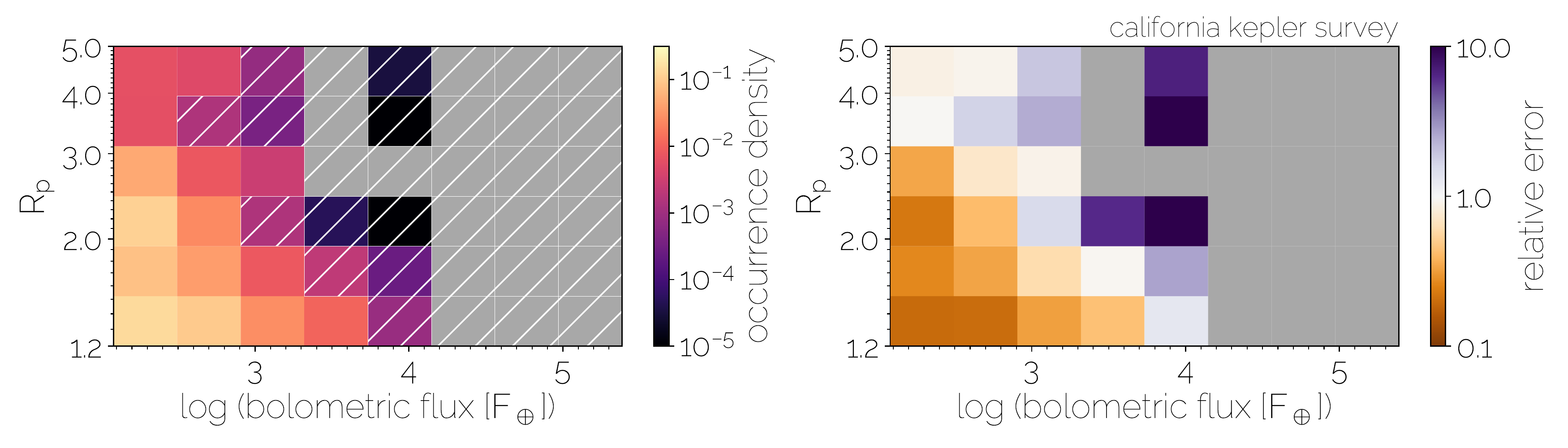}
\includegraphics[width = \textwidth]{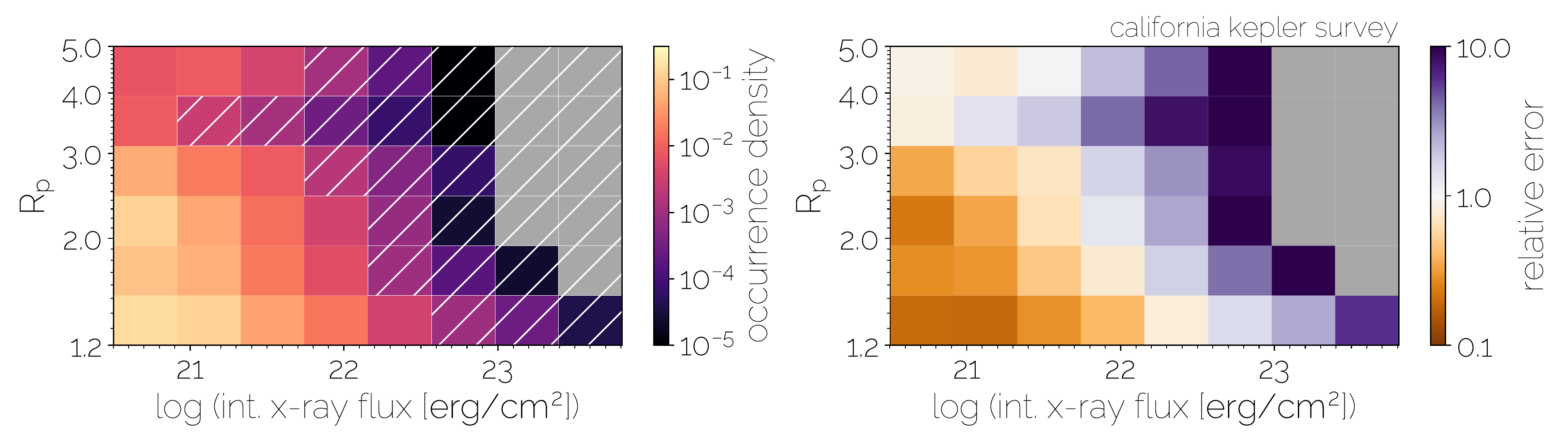}
\caption{Occurrence densities for the subsample of the California Kepler Survey, specified on a grid for which the maximum relative error in a grid cell is $\sim 10$. The photoevaporation desert is shown as the hatched region and defined as an area consistent with zero planets at the 2$\sigma$ level. Plots are for a) the present-day bolometric flux and c) the lifetime-integrated flux parameter spaces. The flux bounds of plots a) and c) are set to be
equivalent for a solar mass star of age 5 Gyr. Plots c) and d) indicate the relative errors for the respective plots.}
\label{fig7}
\end{figure*}

\subsection{Analyzing the photoevaporation desert in two dimensions}
\label{sub:res_cks_des}

We now define the location of the photoevaporation desert, a region statistically consistent with no sub-Neptune sized planets at high irradation, possibly as a result of atmospheric photoevaporation stripping these planets down to rocky cores of size $<$ 1.4 $R_p$. In \textbf{Fig. 7}, we use the methodology described in section \ref{subsub:define_desert} to define the location of the photoevaporation desert in the CKS data. The hatched region in \textbf{Fig. 7a} defines a region of total occurrence consistent at the $2\sigma$ level with an occurrence of zero, in the parameter space of present-day bolometric flux and planet radius. This is the parameter space that most previous discussions of the photoevaporation desert in the data, e.g. \citet{Lundkvist2016} and \citet{Fulton2017} have used. In \textbf{Fig. 7c} the hatched region defines the photoevaporation desert (with the same criteria for occurrence) in the parameter space of lifetime-integrated X-ray flux. The bounds for both plots are chosen to be equivalent for a planet orbiting a solar mass star of age 5 Gyr.

In both parameter spaces, the desert boundary has a negative slope in $R_p$ vs flux. In bolometric flux, the highest flux value at which the desert boundary is drawn is 2.09 $\times 10^{3} F_{\oplus}$ and located at a radius of 1.52 -- 1.93 $R_{\oplus}$. The lowest flux at which the desert boundary is drawn 3.12 $\times 10^{2} F_{\oplus}$, which is found at a radius of 3.11 -- 3.94 $R_{\oplus}$. The highest and lowest fluxes for lifetime-integrated X-ray flux are found at the same two radii ranges, with values of 1.43 $\times 10^{22}$ $erg/cm^2$ and 8.23 $10^{20}$ erg/cm$^2$ respectively. Despite the maximum and minimum fluxes of the desert boundaries occurring at the same radii, there are hints of differences in the shape of the desert between the two parameter spaces. In lifetime-integrated X-ray flux, the flux value at which the desert boundary is drawn appears to decrease monotonically as radius increases until a minimum at 3.1 -- 3.9 $R_{\oplus}$, before increasing again. For bolometric flux, this same monotonic behavior is not found, with a concavity in the desert boundary at 2.4 -- 3.1 $R_{\oplus}$. We note, however, that the shape of the drawn desert is sensitive to the errors of planet parameters, and more rigorous sensitivity testing would be required to make statistically significant claims on the differences in desert shape.

\textbf{Fig. 7b} and \textbf{Fig. 7d} show the relative errors in the occurrences of \textbf{Fig. 7a} and \textbf{Fig. 7c} respectively. The errors are smallest at the center of the planet distributions where large numbers of planets are found in each grid cell, and are greatest at the edges of the distribution, including the desert, where grid cells can consist of just several fractional planets.

The radius and flux values that define the desert boundaries are recorded in \textbf{Table 2}. Comparisons of the desert we have found with other data sets will be made in section \ref{sub:desert_discuss}

\begin{longrotatetable}
\begin{deluxetable*}{lcccccc}
\tablecolumns{7}
\tablewidth{0pt}
\tablecaption{Photoevaporation desert boundaries as a function of planet radius.}
\tablehead{\colhead{Data set} & \colhead{\qquad} & \colhead{\qquad} & \colhead{Planet radius range} & \colhead{\qquad} & \colhead{\qquad} & \colhead{\qquad}}
\startdata
\qquad & 1.2 -- 1.52 & 1.52 -- 1.93 & 1.93 -- 2.45 & 2.45 -- 3.11 & 3.11 -- 3.94 & 3.94 -- 5.0 \\
\qquad & \qquad & \qquad & \colhead{Bolometric flux ($F_\oplus$)} & \qquad & \qquad & \qquad \\
California Kepler Survey & 5.41 $\times 10^{3}$ & 2.09 $\times 10^{3}$ & 8.07 $\times 10^{2}$ & 2.09 $\times 10^{3}$ & 3.12 $\times 10^{2}$ & 8.07 $\times 10^{2}$ \\
\qquad & \qquad & \qquad & \colhead{Lifetime-integrated X-ray flux (erg/cm$^2$)} & \qquad & \qquad & \qquad \\
California Kepler Survey & 3.70 $\times 10^{22}$ & 1.43 $\times 10^{22}$ & 1.43 $\times 10^{22}$ & 5.52 $\times 10^{21}$ & 8.23 $\times 10^{20}$ & 5.52 $\times 10^{21}$ \\
\colhead{Q1-17 DR 25} & \qquad & \qquad & \qquad & \qquad & \qquad & \qquad \\
\qquad M \& K subsample & 1.50 $\times 10^{22}$ & 1.50 $\times 10^{22}$ & 1.50 $\times 10^{22}$ & 5.71 $\times 10^{21}$ & 2.18 $\times 10^{21}$ & 2.18 $\times 10^{21}$ \\
\qquad G subsample & 1.50 $\times 10^{22}$ & 1.50 $\times 10^{22}$ & 5.71 $\times 10^{21}$ & 2.18 $\times 10^{21}$ & 2.18 $\times 10^{21}$ & 2.18 $\times 10^{21}$ \\
\qquad F subsample & 8.33 $\times 10^{20}$ & 8.33 $\times 10^{20}$ & 3.18 $\times 10^{20}$ & 3.18 $\times 10^{20}$ & 3.18 $\times 10^{20}$ & 3.18 $\times 10^{20}$ \\
\enddata
\end{deluxetable*}
\end{longrotatetable}

\subsection{The photoevaporation desert in one dimension: Cumulative distribution functions}
We use cumulative distribution functions (CDFs) over the flux dimension, beginning with the highest fluxes, to examine the location of the photoevaporation desert, which is represented in the higher-flux portion of the CDF, as well as the shape of the photoevaporation desert, which is represented in the slope of the CDF. The shape of the photoevaporation desert is also represented in the probability density functions (PDFs) that we generate, which show variations in the probability density of planets as a function of flux. In \textbf{Fig. 8a}, we compare cumulative distribution functions (CDFs) of the gaseous and rocky planets in the CKS data set in the parameter space of lifetime-integrated X-ray flux. We have defined rocky to be planets with radii of 1.2 -- 1.8 $R_{\oplus}$, and gaseous planets to be between 1.8 -- 4 $R_{\oplus}$. The Anderson-Darling (AD) test on the two cumulative distribution functions returns an AD statistic of $9 \times 10^3$. Because this is much greater than the 1\% critical value of 3.75, we consider the flux-dependent behavior of the rocky and gaseous planets to be different with 99\% confidence.

In addition, both the CDF and PDF (\textbf{Fig. 8b}) show that the distribution of rocky planets is shifted to higher fluxes than the gaseous planets. The fact that there is an underdensity of high-flux gaseous planets, relative to rocky planets, is suggestive of a process shaping the gaseous planet population which is not affecting the rocky planets---namely photoevaporation.

\begin{figure*}
\gridline{\fig{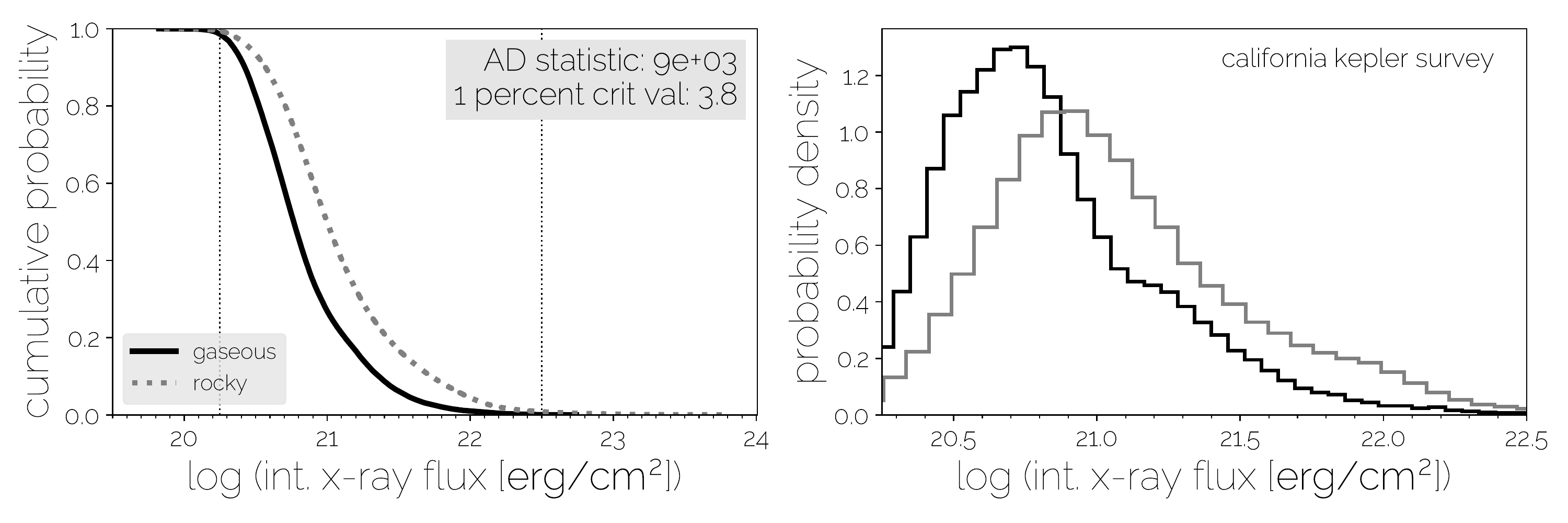}{\textwidth}{a) \hspace{70pt} b)}}
\caption{a) A comparison of the cumulative distribution functions (CDF) for the
rocky super-Earths and gaseous sub-Neptunes (which we define as between 1.2 -- 1.8 $R_\oplus$, and 1.8 -- 4 $R_\oplus$ respectively) in the subsample of the California Kepler
Survey. The Anderson-Darling statistic comparing the two CDFs is shown. b) The probability distribution function (PDF), which indicates the onset of the desert location as well as the sharpness of the transition from the photoevaporation desert to the center of the planet distribution.}
\label{fig8}
\end{figure*}

\begin{figure*}
\gridline{\fig{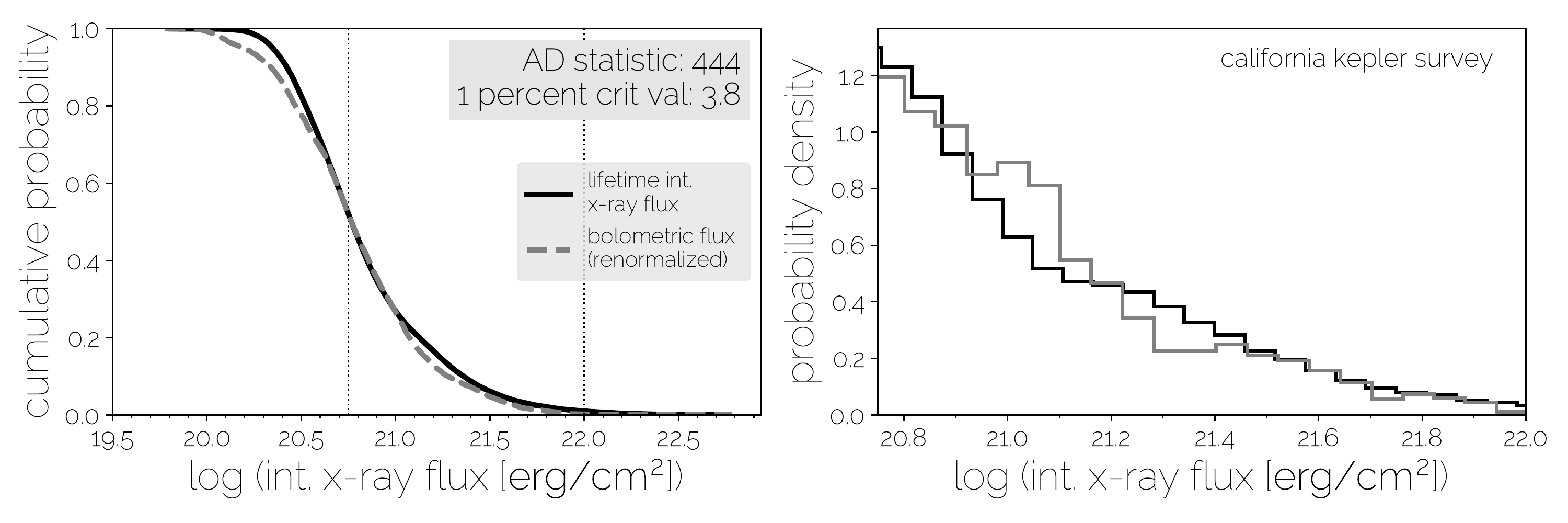}{\textwidth}{a) \hspace{70pt} b)}}
\caption{a) The CDF of all sub-Neptune sized planets in the California Kepler Survey. We compare the CDFs of the planets in the parameter spaces of present-day bolometric flux and lifetime-integrated X-ray flux, and the AD statistic for this comparison is shown. For this comparison, the bolometric flux data are renormalized to the lifetime-integrated X-ray flux data such that they have identical values at the 10th and 90th percentile of flux. b) The PDF, containing the same information as described in \textbf{Fig. 8b}.}
label{fig9}
\end{figure*}

\begin{figure*}
\gridline{\fig{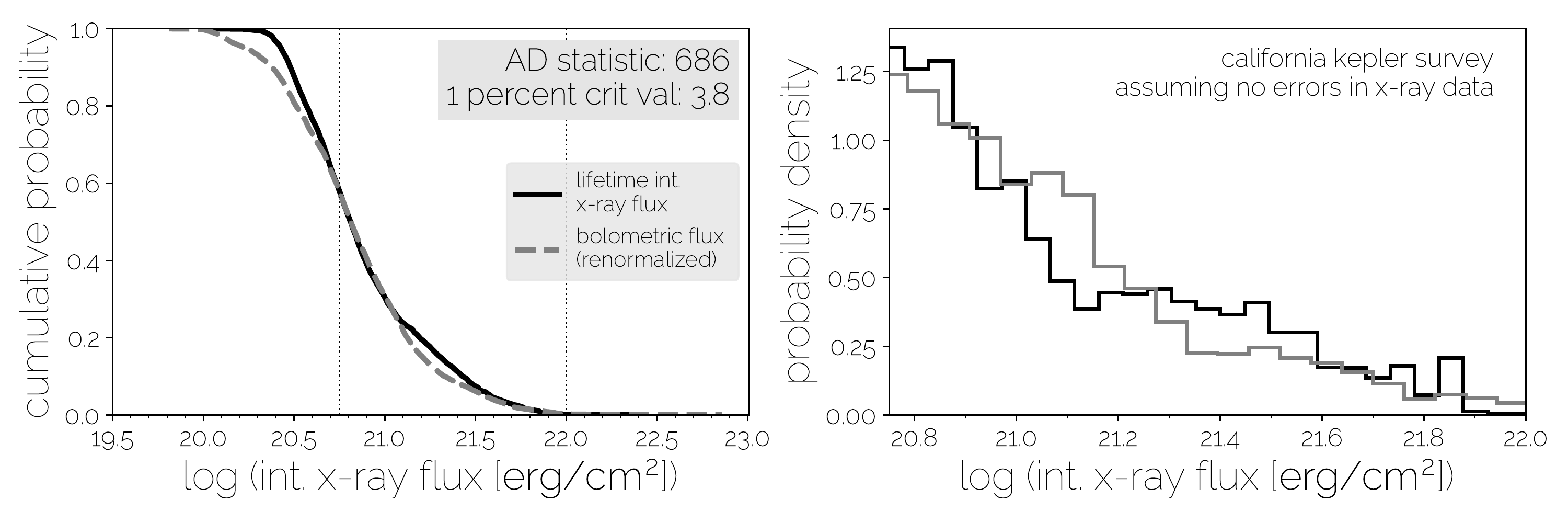}{\textwidth}{a) \hspace{70pt} b)}}
\caption{a) The CDFs shown in Fig. 9, with the exception that for the lifetime-integrated X-ray calculation, the X-ray luminosities as a function of age and stellar mass are assumed to be known exactly. b) The PDF, containing the same information as described in \textbf{Fig. 8b}.}
\label{fig10}
\end{figure*}

\begin{figure*}
\gridline{\fig{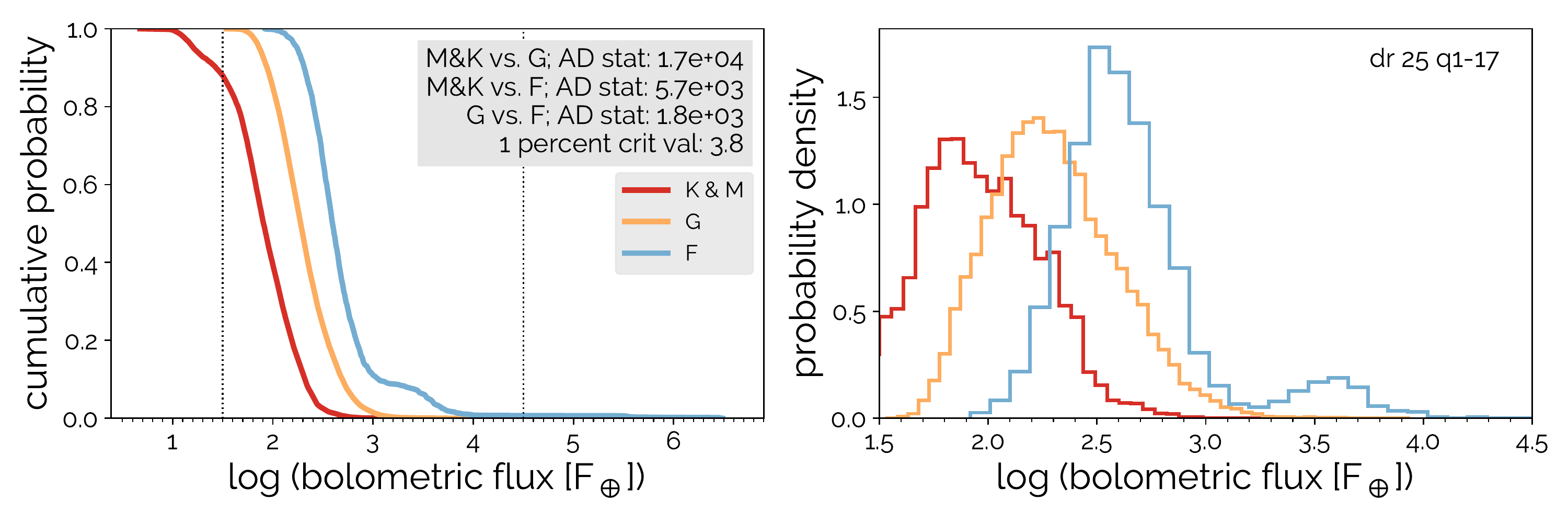}{\textwidth}{a) \hspace{70pt} b)}}
\caption{a) The CDFs of all sub-Neptune sized planets as a function of stellar type for the Q1-17 DR 25 subsample in the bolometric flux parameter space, as well as the AD statistics comparing each CDF. b) The PDF, containing the same information as described in \textbf{Fig. 8b}.}
\label{fig11}
\end{figure*}

\begin{figure*}
\gridline{\fig{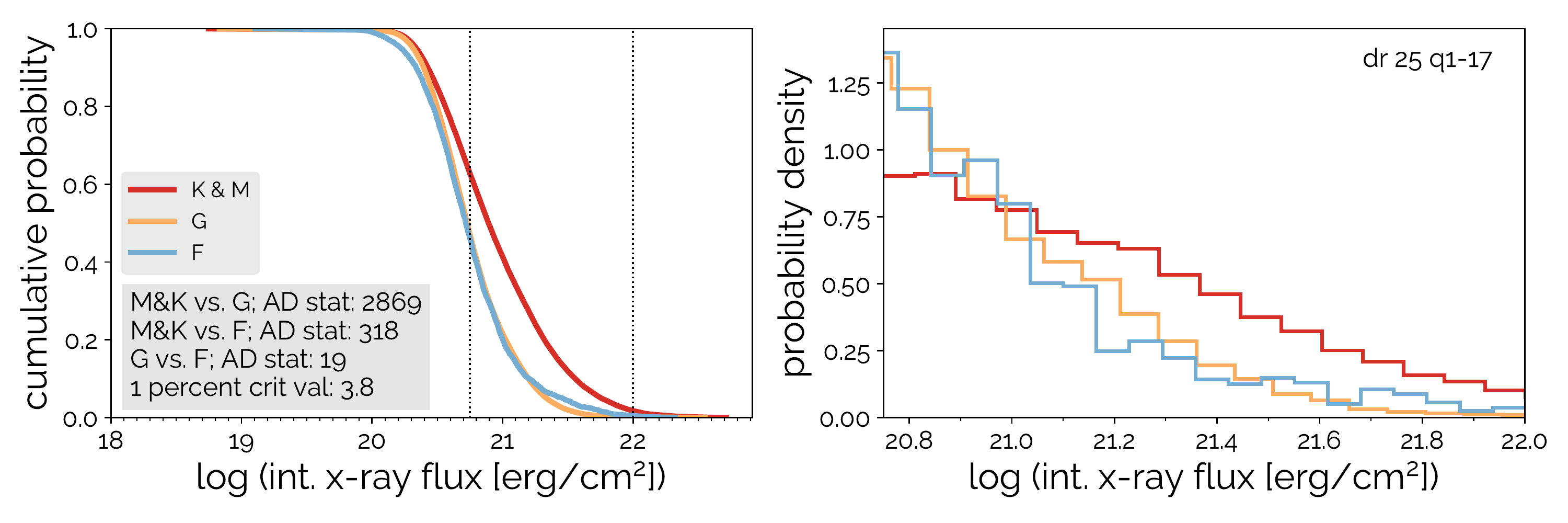}{\textwidth}{a) \hspace{70pt} b)}}
\caption{a) The CDFs of all sub-Neptune sized planets as a function of stellar type for the Q1-17 DR 25 subsample in the lifetime-integrated X-ray flux parameter space, as well as the AD statistics comparing each CDF. b) The PDF, containing the same information as described in \textbf{Fig. 8b}.}
\label{fig12}
\end{figure*}

In \textbf{Fig. 9a} the CDFs for our CKS sample are compared in the parameter spaces of bolometric flux and lifetime-integrated X-ray flux. To do this comparison, we renormalize the bolometric flux distribution to the 10$^{th}$ and 90$^{th}$ percentiles of the lifetime-integrated X-ray flux distribution (see section \ref{subsub:cmd}). The AD statistic of 444 is again larger than the 1\% critical value of 3.75, indicating with 99\% confidence that the distribution of planets in bolometric flux and lifetime-integrated X-ray flux are different. For the PDF in \textbf{Fig. 9b}, one curve does not exhibit consistently steeper slopes than the other, with the exception of some small underdensities and overdensities in the bolometric flux distribution. By casting the photoevaporation desert in a parameter space directly tied to the physical process of photoevaporation, lifetime-integrated X-ray flux, one might expect the desert to be more sharply defined when compared to the bolometric flux parameter space. Nevertheless, the CKS sample consists of largely Sun-like stars, and accounting for differences in stellar type is the biggest reason why lifetime integrated X-ray flux should produce a sharper desert edge compared to bolometric flux---an effect that would not be dramatic in this sample. In section \ref{sub:res_stlr_des}, we will examine the CDFs for the DR 25 Q1-17 dataset, which encompass a wider variety of stellar types.

We consider the possibility that a steeper CDF for the lifetime-integrated X-ray flux is not apparent due to the errors in the X-ray observations. \textbf{Fig. 10} shows the same CDF comparison of \textbf{Fig. 9}, with the exception that the X-ray luminosity is assumed to be known exactly as a function of stellar mass and age (e.g. \textbf{Fig. 2}, but with relative errors of 0 across the entire parameter space). Although the AD statistic for the two distributions again indicates that the distributions are distinct, once again neither the slope of the CDF for lifetime-integrated X-ray flux nor the slope of the CDF for bolometric flux is consistently greater than the other. Assuming there are no systematic errors in the X-ray observations of \citet{Jackson2012} and \citet{Shkolnik2014}, which seem unlikely given that their relations are largely in agreement with theoretical considerations, the steepness of the slopes for the sub-Neptune planet distribution in lifetime-integrated X-ray flux space and bolometric flux space appear to be largely indistinguishable.

\subsection{The desert as a function of stellar type}
\label{sub:res_stlr_des}

We now examine differences in the photoevaporation desert as a function of stellar type. Examining the desert for a variety of stellar types is insightful because bolometric flux varies by close to 2 orders of magnitude among FGK stars, but the lifetime x-ray luminosity is roughly flat with stellar mass for $M_*$ $>$ 0.6 $M_{\astrosun}$.

\textbf{Fig. 11} compares the CDFs for the different stellar types as a function of bolometric flux, using the subsample of the DR 25 Q1-17 dataset, chosen to minimize possible effects from incompleteness when compared with other sources of error (see section \ref{subsub:complete}). This sample is using stellar properties corrected from the original DR 25 catalog, i.e. the DR 25 Supplemental Stellar data. In the CDFs (\textbf{Fig. 11a}) as well as the PDFs (\textbf{Fig. 11b}), it is clear that the distributions of the planets around different stellar types peak at different fluxes. The AD statistics are all $> 10^3$, indicating that all distributions are significantly distinct from one another.

\textbf{Fig. 12} compares the CDFs for the different stellar types using the same sample, but in the lifetime-integrated X-ray flux parameter space. The similarity among the distributions in lifetime X-ray flux compared to bolometric flux \textbf{Fig.11} is immediately apparent. All three distributions peak around $10^{20.7}$ erg/cm$^2$. Although the AD statistic indicates that all distributions are still distinct from one another, the AD statistics in lifetime X-ray flux are 1 -- 2 dex lower than in bolometric flux. This suggests that the primary reason for differences in the distributions as a function of stellar type are a result of bolometric flux variations as well as variations in the $L_X/L_{bol}$ ratio as a function of stellar type. When the planet distribution is cast into lifetime-integrated X-ray flux, the parameter physically tied to the evaporation process, the bulk of the variability among stellar types is corrected for. The differences that remain are possibly due to non-steady-state X-ray emissions (flares and coronal mass ejections) that are not measured in the X-ray data, as well as other secondary processes. Another possiblity is variations in data precision as a function of stellar type---wherein uncertainties for late stellar types typically have slightly larger average uncertainties.

The difference between the M \& K and G type CDFs (AD statistic = 1251) is found to be considerably greater than between the M \& K and A \& F samples (AD stat. = 33), or between the G and A \& F samples (AD stat. = 88). In comparing the PDFs (\textbf{Fig. 12b}), the larger difference between the CDFs for the M \& K and G type stars is apparent in a sharper transition for the G-type stars between the desert and the center of the planet distribution---starting out at high fluxes with the smallest planet probability density, until transitioning to the highest planet probability density at the median flux.

\begin{figure*}
\includegraphics[width = \textwidth]{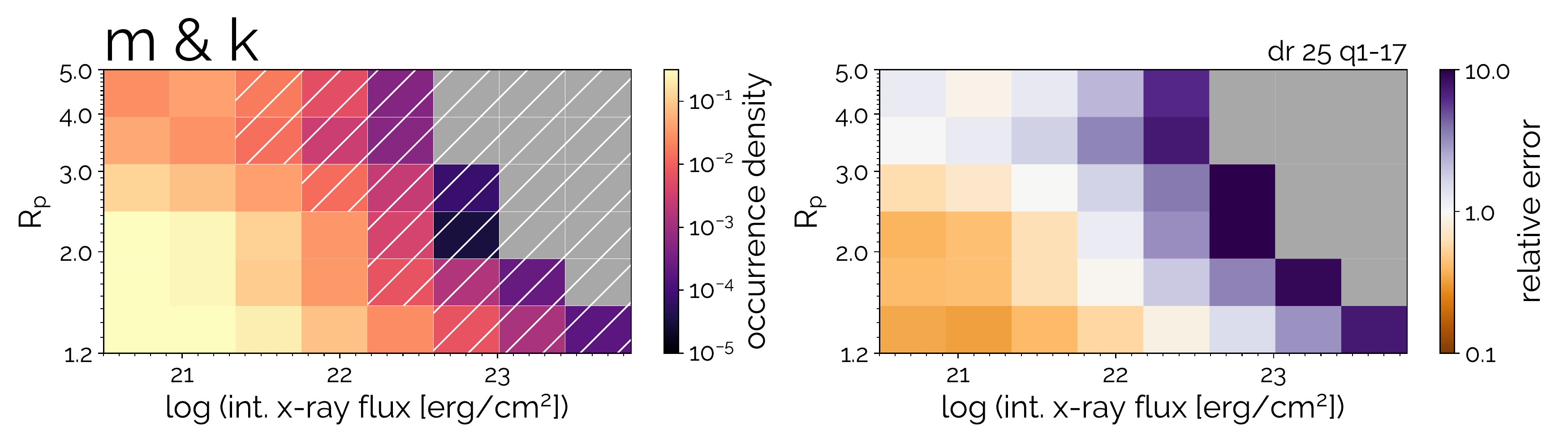}
\includegraphics[width = \textwidth]{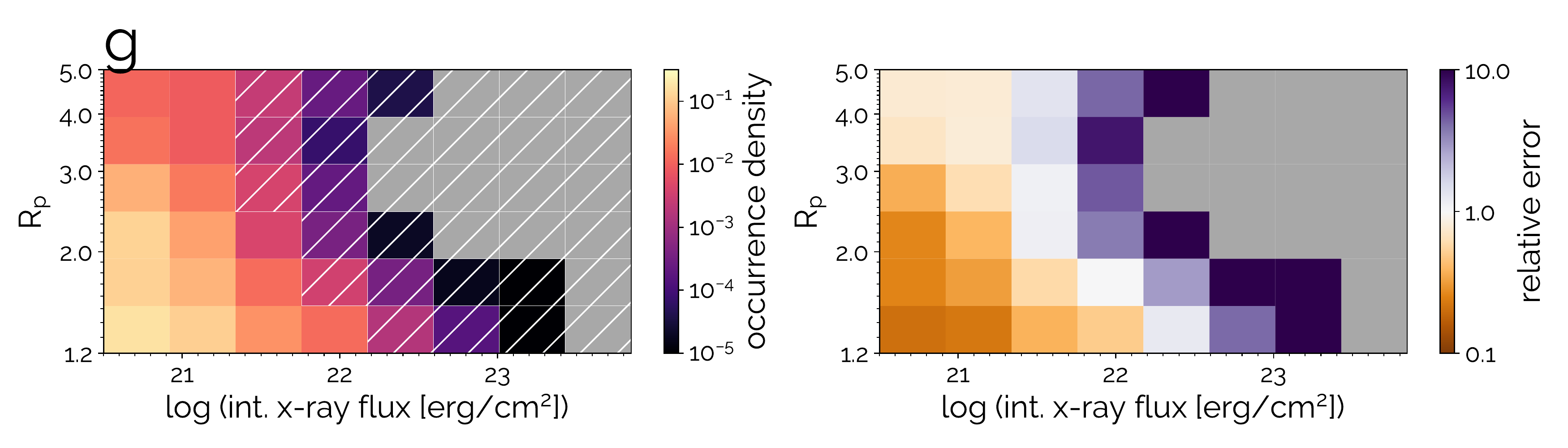}
\includegraphics[width = \textwidth]{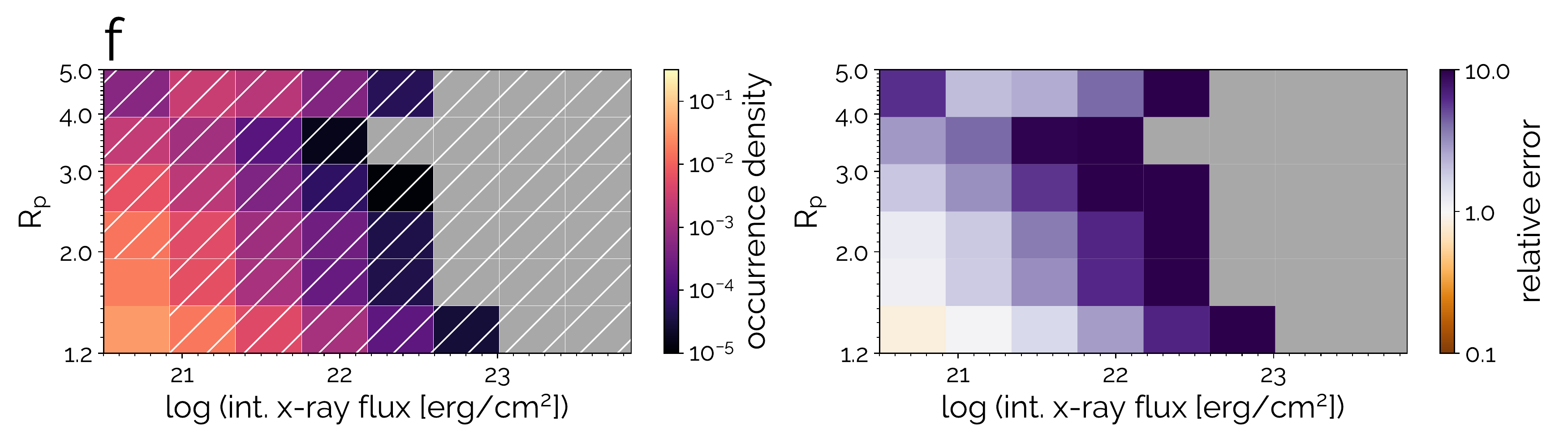}
\caption{Changes in planet occurrence and the shape of the photoevaporation
desert as a function of stellar type for the Q1-17 DR 25 subsample. a) M \& K type stars c) G type stars e) F type stars. The relative errors for the above subsamples are shown in subplots b), d) and f) respectively. The plotting conventions are the same as for \textbf{Fig. 7}}
\label{fig13}
\end{figure*}

\textbf{Fig. 13} compares the stellar types in the radius dimension in addition to the flux dimension. The consistency in the location of the desert among the sample of M \& K type stars (\textbf{Fig. 13a}) and G type stars (\textbf{Fig. 13c}) in lifetime-integrated X-ray flux is visible.  The desert drawn for the F type stars is significantly larger than for either the M \& K or G samples. However, we note that the number of planets in the F sample is considerably smaller than in the other samples (fractional planets equivalent to a total of 6.2 planets, compared with 52.9 for the M \& K sample, and 97.4 for the G sample, see \textbf{Table 1}). For this reason, the errors in the desert boundaries drawn for the F sample will be much larger than for the other samples---this could very possibly be the reason for the difference in the shape of the drawn desert region for the F-type stars in two dimensions.

\begin{figure*}
\gridline{\fig{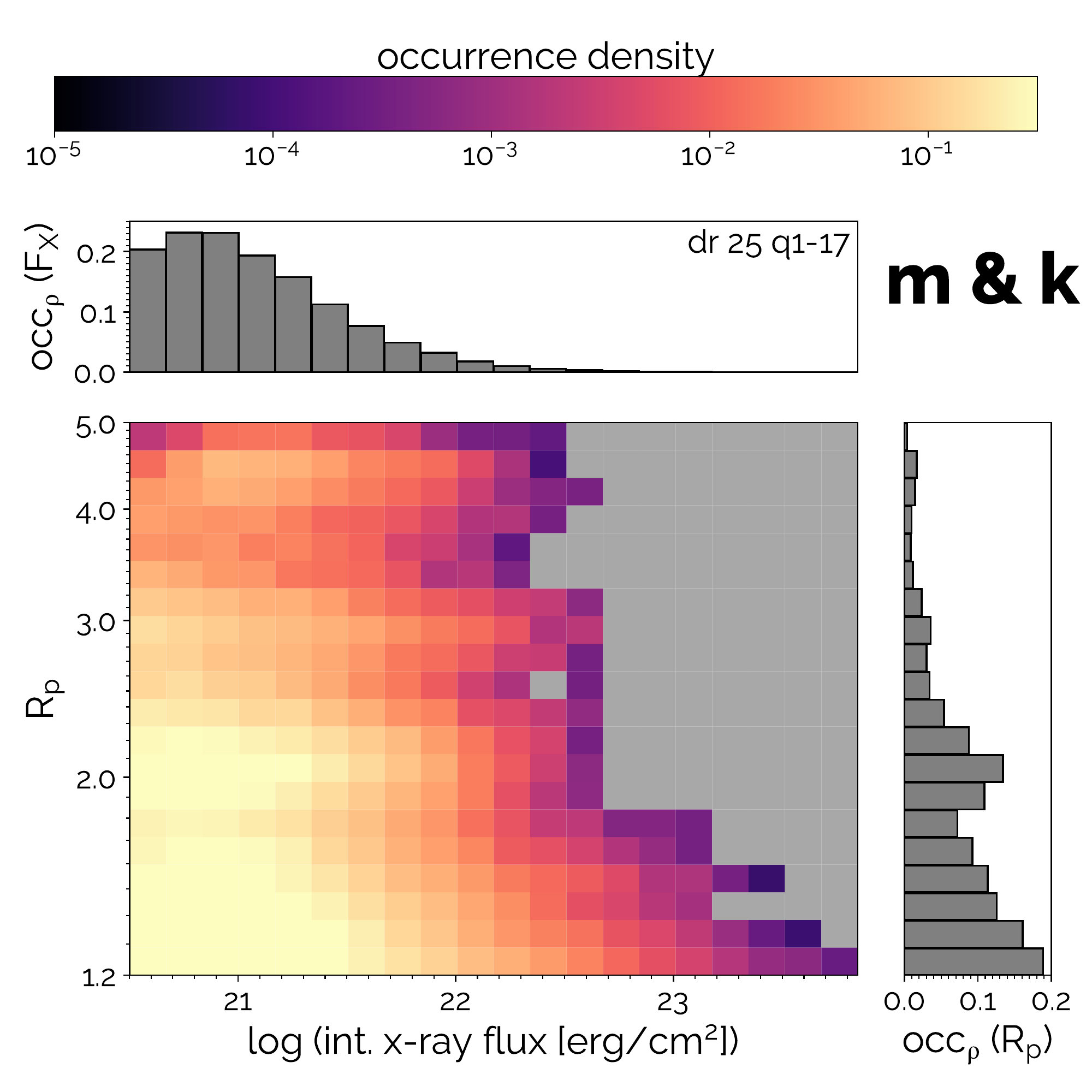}{0.33\textwidth}{a)}
            \fig{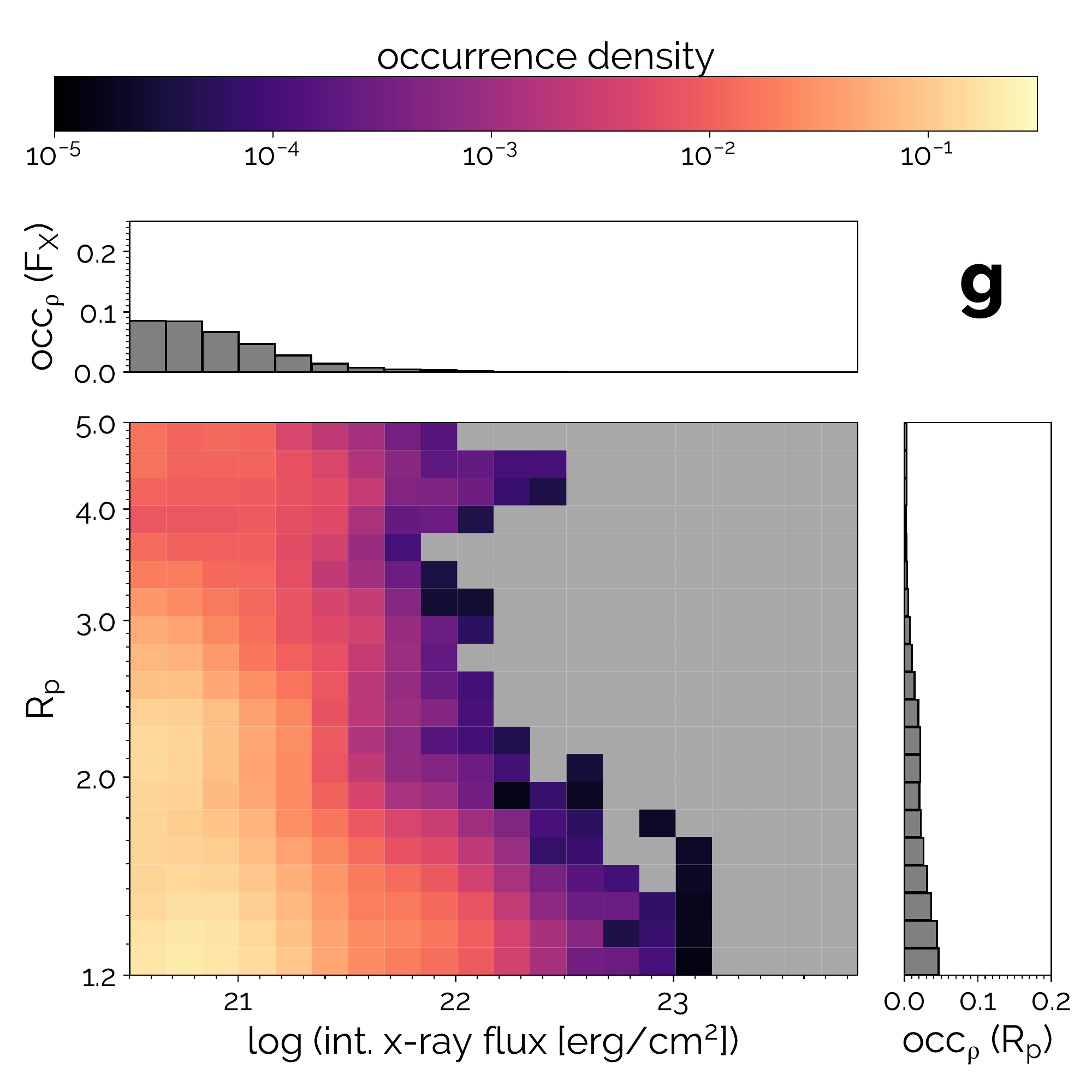}{0.33\textwidth}{b)}
                \fig{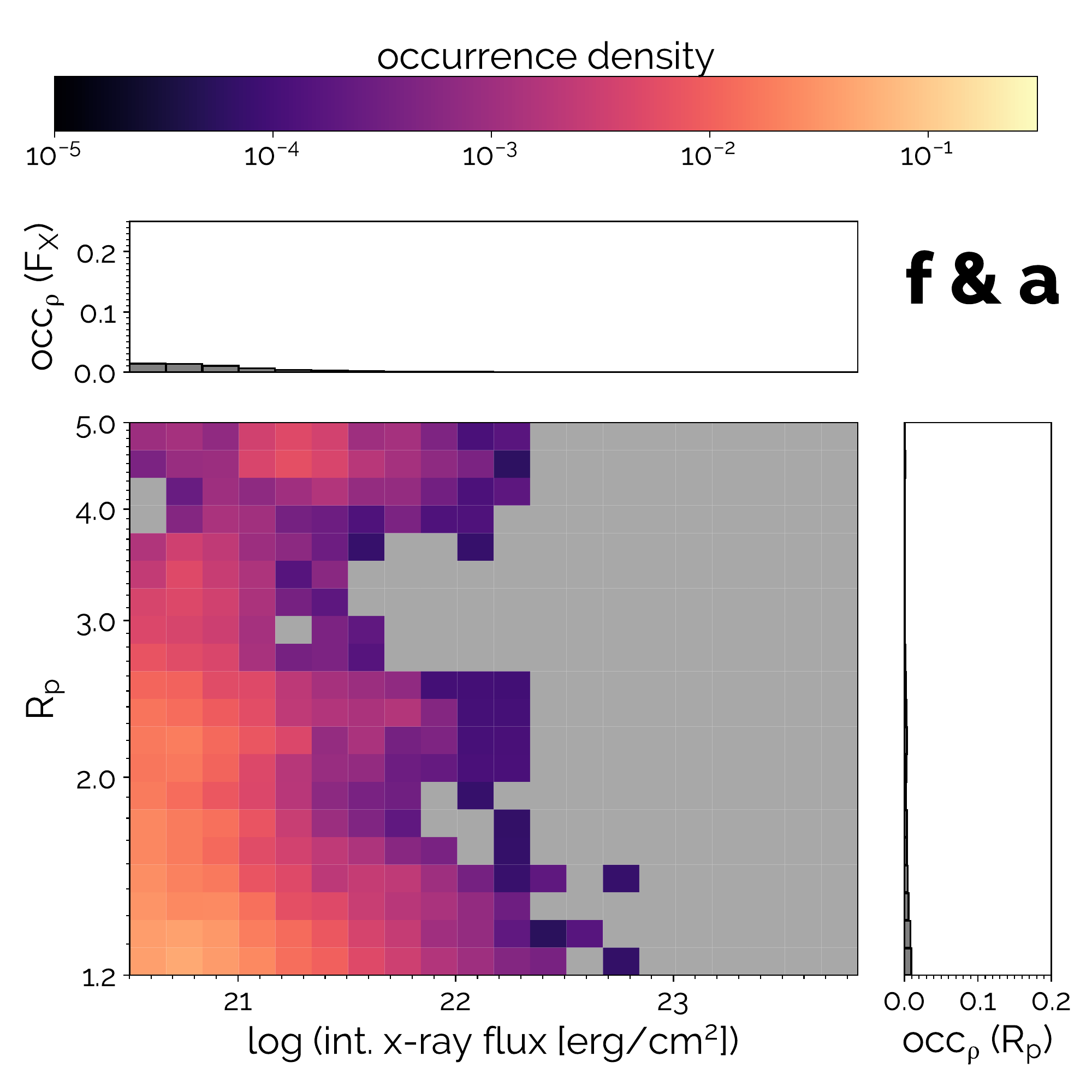}{0.33\textwidth}{c)}}
\caption{Summed 1D occurrences over planet radius and lifetime-integrated X-ray flux, as a function of stellar type. a) M \& K type stars b) G type stars c) F type stars. The plotting conventions are the same as for \textbf{Fig. 6}}
\label{fig14}
\end{figure*}

\subsection{The photoevaporation valley as a function of stellar type}
\label{sub:res_valley}

We now examine the photoevaporation valley, the gap in the \textit{Kepler} planet data at 1.5 -- 2 $R_{\oplus}$ first reported by \citealt{Fulton2017}, using the same stellar subsamples of section \ref{sub:res_stlr_des}. The photoevaporation valley occupies a distinct parameter space from the photoevaporation desert, and for a visual comparison the reader is referred to \textit{Fig. 10} of \citet{Fulton2017}. We now look at the samples at a finer grid resolution in \textbf{Fig. 14} so as to investigate in detail the occurrences summed over the radius and integrated X-ray flux dimensions. Although the parameter space is now oversampled in two dimensions, the relative errors for the occurrences for the M \& K and G-type stars in one dimension (over radius and flux space, visible as the 1D histograms on the axes of the plots) are still less than our criteria of $< 1$ dex. The valley is clearly visible in the right panel of \textbf{Fig. 14a}, which consists of the M \& K type star subsample, as a minimum in occurrence at $R_p = 1.6 R_{\oplus}$. This represents an advance to studies of the photoevaporation valley in a suggestion of its presence for later stars; the original discovery of \citealt{Fulton2017} reported the valley for solar-type stars hotter than $\sim 4600$ K.

\begin{figure}
\includegraphics[width=0.5\textwidth]{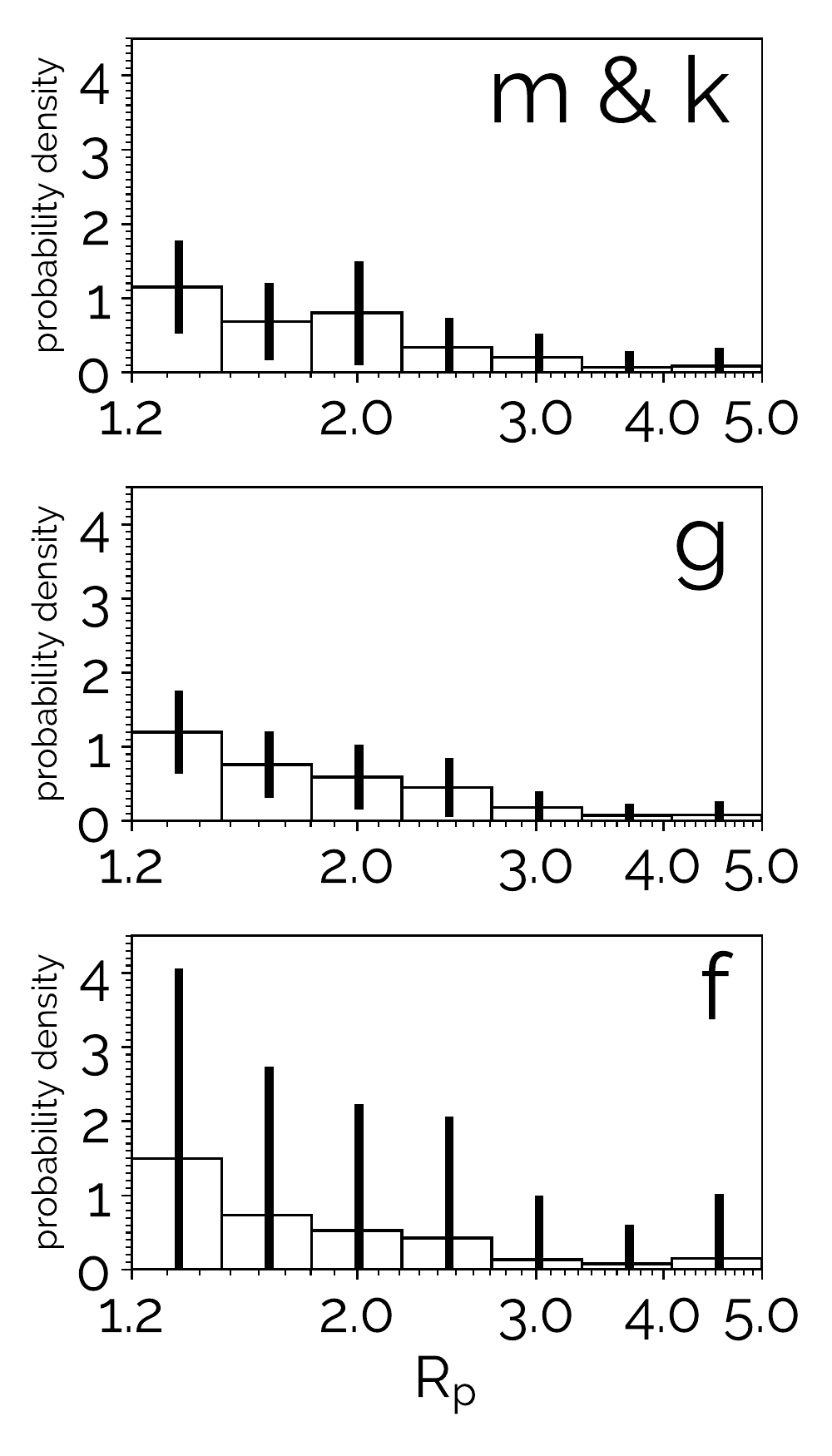}
\caption{Histograms comparing the relative occurrence densities of planets as a function of radius among stellar types. The lines denote 1 $\sigma$ error bars.}
\label{fig15}
\end{figure}

In \textbf{Fig. 15} we show these same occurrence densities marginalized as a function of radius, with the distributions now normalized to probability densities, facilitating comparisons among stellar types, as well as with 1 $\sigma$ error bars (calculated using equation \ref{occ_std}). The M \& K radius distribution is indicative of a valley at $R_p = 1.6 \oplus$ at the 1 $\sigma$ level. The valley is not apparent in the G sample, and cannot be reported on at the 1 $\sigma$ level for the F type stars.

\section{Discussion}

\subsection{Comparison of our photoevaporation desert with previous studies}
\label{sub:desert_discuss}

The desert we defined in section \ref{sub:res_cks_des} is in good agreement with previous theoretical studies. The slope of our desert boundary drawn in \textbf{Fig. 7b} is qualitatively consistent with the slope of the theoretical curve presented in \textit{Fig. 7} of \citealt{Owen2013}, with the intercept of our boundary in lifetime-integrated X-ray flux offset by roughly a half order of magnitude to higher fluxes. Although \citealt{Owen2013} made a visual comparison of their theoretical curves with a previous \textit{Kepler} table, our having defined a desert boundary directly from the data using statistical considerations allows for direct quantitative comparisons with evaporation models.

In the parameter space of bolometric flux, the desert region identified by \citet{Lundkvist2016} can be considered a subset of the desert we have drawn. However, rather than try to define a region completely absent of planets, we have shown that a much larger portion of the parameter space spanning lifetime-integrated X-ray fluxes from 8.23 $\times 10^{20}$ to $10^{24}$ erg/cm$^2$ (see \textbf{Table 2} for tabulations of the desert boundary) can be considered consistent with zero planets at the 2 $\sigma$ level.

\subsection{Evidence that the gaseous planets are sculpted by photoevaporation, and a depletion in gaseous planets at high lifetime integrated X-ray fluxes}

Our finding of the center of the planet distribution for gaseous planets ($1.8 < R_{\oplus} < 4$) occuring at lower lifetime-integrated X-ray fluxes compared to rocky planets ($1.2< R_{\oplus} < 1.8$) may be indicative of the distribution of gaseous planets as a function of lifetime-integrated X-ray flux being shaped by the photoevaporation process. Similarly, it suggests a depletion in gaseous compared to rocky planets at higher fluxes.

We find that the utility of examining the photoevaporation desert in the lifetime-integrated X-ray flux parameter space is most pronounced when looking at a sample of planets orbiting a large variety of stellar types. This is apparent in the transition of the stellar subsamples from bolometric flux space in \textbf{Fig. 11}, where the distributions peak at different fluxes, to lifetime X-ray flux where the stellar subsamples pile up \textbf{Fig. 12}. The 1 -- 2 dex drop in the AD statistics after this transition are indicative of variations in bolometric flux and the $L_X/L_{bol}$ ratio as a function of stellar type being the number one source of variation in the planet distributions of sub-Neptunes. When these variations are corrected for and the sub-Neptunes are examined in lifetime X-ray flux space, we are left with a sample more resembling a single distribution. We consider these photoevaporative considerations to be the primary candidate for shaping our observed planet distributions as a function of stellar types, as dynamical considersations which also vary as a function of stellar type (e.g. tidal loss) take place at considerably shorter orbital periods (e.g. several hours for Sun-like stars) than even the tails of our planet distributions.

\subsection{Differences in the desert among stellar types: Flares and coronal mass ejections as well as predictions for future surveys of M dwarfs}

After correcting for the bolometric variations among stellar types using lifetime-integrated steady-state X-ray fluxes, whatever variations that remain among stellar types could possibly be a result of secondary X-ray emissions from flares and coronal mass ejections, processes that are also expected to vary as a function of stellar type. Although obvious physical signatures of these processes are not apparent in our results, we suggest that our approach of accounting for steady-state X-ray emissions when examining the photoevaporation desert, coupled with higher precision measurements of late-type stars using a methodology similar to that of the California Kepler Survey, could place observational constraints on the importance of these secondary X-ray emission sources on the photoevaporation process. Furthermore, despite the variety of stellar types present in the Q1-17 DR 25 \textit{Kepler} catalog, the lowest mass stars are still early M dwarfs, whereas massive increases in flaring activity begin with mid M-dwarfs later than M3. Current and future surveys will discover many more planets around mid M dwarfs. These include the \textit{K2} mission, the \textit{Transiting Exoplanet Survey Satellite} (\textit{TESS}), the \textit{Planetary Transits and Oscillations of stars} (\textit{PLATO}) mission, as well as the Characterizing Exoplanets Satellite (\textit{CHEOPS}). The addition of the planets discovered by these missions, in combination with our method of correcting for steady-state X-ray emissions, could evaluate the significance of flaring in contributing to the photoevaporation process.

\subsection{The photoevaporation valley for late type stars}
\label{sub:valley_discuss}

Our findings are suggestive at the $1 \sigma$ level of a photoevaporation valley for the M \& K dwarfs, centered at $\sim 1.6 R_{\oplus}$. Thus the valley may not be unique to Sun-type stars, and higher precision measurements of late-type stars should allow for rigorous claims on its dependence on stellar type. Our suggestion that the photoevaporation valley continues for planets around for M \& K dwarfs motivates such follow-up measurements, which in turn may allow for interpretation of the origins for the photoevaporation valley---whether it be planet core luminosity driving atmospheric loss \citep{Ginzburg2017} or the same photoevaporation process responsible for the photoevaporation desert \citep{Owen2017}.

\section{Summary}

We have examined how lifetime-integrated X-ray flux varies as a function of stellar mass and age. This knowledge was used to cast the \textit{Kepler} dataset into the lifetime-integrated X-ray flux parameter space where statistical constraints were placed on the photoevaporation desert, as well as the photoevaporation valley for late type stars. The key results of this study are summarized below.

(i) Lifetime-integrated X-ray luminosity decreases with stellar mass, and almost all of this drop happens for M-dwarfs $<$ 0.6 $M_{\astrosun}$. Above this mass, the lifetime X-ray luminosity is roughly flat with stellar mass since increases in the saturation level and lifetime

(ii) For a given present day insolation, planets around lower mass stars experience relatively more X-ray flux over their lifetimes ($\sim$100 $\times$ more for a planet orbiting a 0.3 $M_{\astrosun}$ star versus 1.2 $M_{\astrosun}$).

(iii) We have defined a photoevaporation desert, a region consistent with a planet occurrence of zero at the 2 $\sigma$ level. The highest lifetime X-ray flux at which the desert boundary is drawn is 1.43 $\times 10^{22}$ $erg/cm^2$ and located at a radius of 1.52 -- 1.93 $R_{\oplus}$. The lowest lifetime X-ray flux at which the desert boundary is drawn is 8.23 $\times 10^{20}$ erg/cm$^2$, which is found at a radius of 3.11 -- 3.94 $R_{\oplus}$.

(iv) We find that the flux-dependent behavior of rocky and gaseous planets are different with 99\% confidence. We observe that the the distribution of rocky planets is shifted to higher fluxes than the gaseous planets, suggestive of a process shaping the highly irradiated gaseous planet population which is not affecting the rocky planets (which we interpet as photoevaporation).

(v) The utility of examining the photoevaporation desert in the lifetime-integrated X-ray flux parameter space is most pronounced when looking at a sample of planets orbiting a large variety of stellar types. The probability density functions of the sub-Neptunes peak at different bolometric fluxes for different stellar types. Correcting for steady-state stellar X-ray fluxes by casting the sub-Neptune sample in lifetime-integrated X-ray flux space, the planets around all stellar types more closely resemble a single distribution.

(vi) Our findings are suggestive at the $1 \sigma$ level of a photoevaporation valley for the M \& K dwarfs, centered at $\sim 1.6 R_{\oplus}$.

We have not only place improved observational constraints on the location of the photoevaporation desert, but have also developed a methodology to correct for how steady-state X-ray emissions shape the sub-Neptune population as a function of stellar type. We suggest that this technique, combined with the greatly increased sample size of planets around low-mass stars that will be made available by current and future surveys, will be valuable in isolating the effects that secondary X-ray emission sources such as flares and coronal mass ejections have on the photoevaporative process.

\section*{Data Access}
Machine readable versions of the integrated and instantaneous X-ray luminosity as a function of stellar mass and age (shown in \ref{fig2}a and \ref{fig2}c respectively) will be made available on the corresponding author's website, \url{https://www.mcdonaldastro.com}. GDM welcomes any and all inquiries regarding data availability, please do not hesitate to reach out.

\section*{Acknowledgements}
GDM would like to thank all the readers of this paper who have been patient for its upload to \textit{arXiv}, as well as for the hosting of data products. During revision and initial publication, GDM had left academia for ~2 years, and upon returning is working to improve accessibility to his publications and data products.

GDM acknowledges funding from the Department of Defense (DoD) through the National Defense Science \& Engineering Graduate Fellowship (NDSEG) Program. This work was initiated at the 2016 Kavli Summer Astrophysics Program at UC Santa Cruz, and the authors would like to thank Pascale Garaud and Jonathan Fortney for their work in organizing the program. The authors gratefully acknowledge Alan Jackson, Evgenya Shkolnik and Erik Petigura for their providing data from previous publications without which this study could not have been completed. We would also like to acknowledge James Owen, Angie Wolfgang and Ruth Murray-Clay for their input and helpful discussions. EDL is thankful for support from GSFC Sellers Exoplanet Environments Collaboration (SEEC), which is funded by the NASA Planetary Science Divisions Internal Scientist Funding Model. GDM would like to thank Joshua M\'{e}ndez and Sven Simon for their support and encouragement while this research was carried out. GDM would also like to thank James Wray for his supporting my pursuit of this project.
\bibliographystyle{aasjournal}
\bibliography{library.bib}

\listofchanges
\end{document}